\RequirePackage{fix-cm}
\documentclass{svjour3}                     
\smartqed  

\usepackage{algorithmicx}
\usepackage{grffile}
\usepackage{rotating}
\usepackage{array}
\usepackage{lineno}
\usepackage{booktabs} 
\usepackage{graphicx}
\usepackage{epstopdf,epsfig,subfloat,subfig}

\usepackage{amsfonts}
\usepackage{bm,amsmath}
\usepackage{amssymb,color,mathtools,makecell}
\usepackage{multirow}
\usepackage[hyperfootnotes=true]{hyperref}
\hypersetup{
	colorlinks=blue,
	citecolor=Violet,
	linkcolor=Red,
	urlcolor=Blue}

\newcommand{\Kn}{\text{Kn}}
\newcommand{\Ma}{\text{Ma}}
 \journalname{Advances in Aerodynamics}

\begin{document}

\title{Multiscale simulation of rarefied gas dynamics via direct intermittent GSIS-DSMC coupling}

\titlerunning{direct intermittent GSIS-DSMC solver (DIG)}       

\author{Liyan Luo \and Lei Wu}
\authorrunning{Luo, Wu} 
\institute{ 
	Department of Mechanics and Aerospace Engineering, Southern University of Science and Technology, Shenzhen 518055, China. \\
	\email{wul@sustech.edu.cn} 
}

\date{Submitted: July 2024 } 

\maketitle

\begin{abstract}
	
	The general synthetic iterative scheme (GSIS) has proven its efficacy in modeling rarefied gas dynamics, where the steady-state solutions are obtained after dozens of iterations of the Boltzmann equation, with minimal numerical dissipation even using large spatial cells. In this paper, the fast convergence and asymptotic-preserving properties of the GSIS are harnessed to remove the limitations of the direct simulation Monte Carlo (DSMC) method. The GSIS, which leverages high-order constitutive relations derived from DSMC, is applied intermittently, which not only rapidly steers the DSMC towards steady state, but also eliminates the requirement that the cell size must be smaller than the molecular mean free path. Several numerical tests have been conducted to validate the accuracy and efficiency of this hybrid GSIS-DSMC approach. 
	
\end{abstract}

\section{Introduction}

The direct simulation Monte Carlo (DSMC) method is widely used to simulate rarefied gas dynamics \cite{Bird1994}. It is efficient for hypersonic flows in the transition and free-molecular flow regimes, but becomes computationally expansive or even prohibitive in the continuum flow regime. That is, in order to reduce the numerical dissipation, the spatial grid size (time step) is chosen to be about one third of the molecular mean free path (mean collision time), resulting in tremendous number of spatial cells and time steps in the continuum flow regime. Therefore, it is urgent to develop numerical schemes with the properties of asymptotic-preserving (reduced to the Navier-Stokes solver in the continuum flow regime, even when the spatial cell size is much larger than the mean free path) and fast convergence (reach the steady state quickly) \cite{Su2020SIAM}. 

Many strategies have been proposed to improve the efficiency of DSMC in the near-continuum flow regime, such as the 
hybrid Navier-Stokes-DSMC method \cite{Wang2003}, the time-relaxed Monte Carlo method~\cite{Pareschi2001}, and the asymptotic-preserving Monte Carlo method~\cite{Ren2014JCP}. 
The first approach needs empirical parameters to determine the region where the Navier-Stokes solver can be applied, while the latter two methods only preserve the Euler asymptotics when the Knudsen number is small. Recently, the time-relaxed Monte Carlo method that accurately preserves the Navier-Stokes equation has been established~\cite{Fei2023JCP}, and several numerical tests are performed to show that it has lower numerical dissipation and higher computational efficiency for multiscale flow simulations. 

In addition to these stochastic methods, in the past decade we have seen great progress in the deterministic simulation of rarefied gas dynamics, but most of the works are based on the simplified gas kinetic models, such as the Bhatnagar-Gross-Krook (BGK)~\cite{bhatnagar1954model} and Shakhov~\cite{Shakhov_S} models. For example, Xu and Huang proposed the unified gas-kinetic scheme~\cite{UGKS2010JCP}, where the streaming and collision are handled simultaneously to reduce the numerical dissipation. Zhu \textit{et al.} developed the implicit unified gas-kinetic scheme, where the use of large time step greatly improves the numerical efficiency~\cite{zhuyajun2016}.
Su \textit{et al.} proposed the general synthetic iterative scheme (GSIS), where the mesoscopic kinetic equation and macroscopic synthetic equations are solved implicitly, and the steady-state solution can be found within dozens of iterations, with negligible numerical dissipation even when large spatial cell size are used~\cite{SuArXiv2019,Su2020SIAM,Liu2024JCP}.

Recently, the community of rarefied gas dynamics has seen the rise of hybrid stochastic-deterministic methodologies. For instance, Degond \textit{et al.} proposed the moment guided Monte Carlo method, where the solution of macroscopic conservative equations are used to guide the evolution of DSMC~\cite{Degond2011}. Liu, Zhu, and Xu introduced the unified gas-kinetic wave-particle methods~\cite{Liu2018arXiv}. In each spatial cell the equilibrium component is addressed by a macroscopic solver, while the non-equilibrium component is tracked by the Monte Carlo approach. By simultaneously handling the streaming and collision processes, this method retains the asymptotic-preserving property of the unified gas-kinetic scheme~\cite{UGKS2010JCP}, yet offers enhanced efficiency for hypersonic flows, as it utilizes simulation particles to represent the molecular velocity space, akin to the DSMC.
Additionally, building upon Fei's research on the asymptotic Navier-Stokes preserving time-relaxed Monte Carlo approach~\cite{Fei2023JCP}, we developed the hybrid GSIS-DSMC method, where the macroscopic synthetic equations exactly derived from the Boltzmann equation are used to expedite the progression of DSMC simulation~\cite{Luo2024arXiv}.

Nevertheless, it is noted that, although these efficient multiscale methods have gained great success in the simulation of single-species rarefied gas dynamics, they are difficult to be extended to simulate rarefied chemical reactions. The reasons are that, i) the DSMC method requires tremendous mathematical skills to develop asymptotic-preserving schemes for flows with multiple relaxation times and complicated physical chemical processes, and ii) for deterministic methods, although asymptotic-preserving and fast convergence can be straightforwardly applied (e.g., see the recent work of GSIS on gas mixtures~\cite{Zeng2024}), it is difficult to construct kinetic models for complicated non-equilibrium chemical reactions.  

Therefore, in this paper, on top of our recent work~\cite{Luo2024arXiv}, we shall develop a simpler numerical framework that possesses the asymptotic-preserving and fast convergence properties, to efficiently and accurately simulate the rarefied gas dynamics. Such a framework is also crafted to facilitate seamless extension to DSMC with chemical reactions. 
For the purposes of clarity and focus, our discussions will be centered on monatomic gas flows.

\section{The intermittent GSIS: Proof of concept}\label{sec_gsis}

The GSIS was initially proposed to solve the Boltzmann equation and its simplified kinetic equations deterministically~\cite{SuArXiv2019}. Its key ingredient is that the mesoscopic kinetic equation and its macroscopic synthetic equation are solved together. While the numerical solution of the kinetic equation provides moment closure to the synthetic equation, the synthetic equation, when solved to the steady state, guides the evolution of velocity distribution functions in the kinetic equation. As a result, both numerical simulations and rigorous mathematical analysis have shown that the GSIS possesses the fast convergence and asymptotic-preserving properties~\cite{Su2020SIAM}, where accurate steady-state solutions are found within dozens of iterations, even when the coarse spatial grid is used. In recently years, the GSIS quickly evolves into a powerful method to study the rarefied gas dynamics~\cite{Su2021CMAME,Liu2024JCP} in practical engineering problems~\cite{Zhang2023arXiv}.

The GSIS has also notably enhanced the convergence of the low-variance (LV) DSMC~\cite{Radtke2009PRE} in the near-continuum flow regime~\cite{Luo2023AiA}. For instance, when the Knudsen number is 0.01, the hybrid GSIS-LVDSMC approach finds the steady-state solution for linearized Poiseuille flow with just 100 spatial cells and $10^4$ evolution steps, contrasting with the 300 cells and $10^5$ steps demanded by the LVDSMC alone. Consequently, the GSIS-LVDSMC reduces the computation time to merely 10 minutes, compared to the 8 hours required by LVDSMC. As the Knudsen number decreases further, the computational advantage of the hybrid GSIS-LVDSMC becomes even more pronounced.

However, the application of GSIS to DSMC is not straightforward, as the linearized BGK equation, whose collision operator is determined by the macroscopic flow quantities such as density, velocity, and temperature, is solved in LVDSMC. Thus, if these macroscopic quantities are obtained by solving the synthetic equation, they can be immediately fed back to LVDSMC to expedite its evolution towards steady state. In contrast, DSMC solves the Boltzmann equation, where the collision operator is determined by the mesoscopic velocity distribution function rather than the macroscopic flow quantities. Therefore, it is imperative to devise a method for integrating macroscopic flow data back into the mesoscopic velocity distribution function (in DSMC it is the distribution of simulation particles). This task, however, is not straightforward.
In our recent paper~\cite{Luo2024arXiv}, this reciprocal feedback is achieved in combination of the asymptotic-preserving time-relaxed Monte Carlo method~\cite{Fei2023JCP}, where the velocity distribution function in the collision process is given analytically by the Wild sum~\cite{Wild1951}. In this method, the Grad-13-type velocity distribution~\cite{Grad1949}, which is determined by the macroscopic quantities from the synthetic equation, corrects the distribution of simulation particles in DSMC in each time step. As a result, fast convergence and asymptotic-preserving are achieved in GSIS-DSMC, resulting a significant improvement of simulation efficiency in the near-continuum flow regime. 

Nevertheless, this GSIS-DSMC coupling is complicated as compared to the original DSMC. First, the collision process in the time-relaxed Monte Carlo method is more time-consuming than the original DSMC. Second, due to the difficulties in the Wild sum with multiple relaxation times, this method, if not impossible, requires tremendous mathematical skills to be extended to flows with chemical reactions. Thus, new methods should be searched to simplify the correction process~\cite{Luo2024arXiv,Fei2023JCP}. 

It is noted that in the deterministic GSIS~\cite{SuArXiv2019,Liu2024JCP} and in the hybrid GSIS-LVDSMC~\cite{Luo2023AiA}, macroscopic quantities obtained from the synthetic equation are used in each iteration/time step. This is not economical in DSMC since the computational time of DSMC in one step is much smaller than solving the synthetic equation for many steps or even to the steady state. In our recent paper~\cite{Luo2024arXiv}, the synthetic equation is solved in every $N$ steps of time-relaxed Monte-Carlo method, but its solution is used to update the simulation particles in every step. 
Out of curiosity, one may ask that what will happen if we solve the synthetic equation and correct the velocity distribution, in every $N$ steps of the kinetic solver? \textbf{We call this method the intermittent GSIS.} Certainly, the value of $N$ cannot be too large, otherwise the benefits of GSIS will be wiped out by the kinetic solver which does not process the asymptotic-preserving property. Also, when the DSMC method is used, the value of $N$ cannot be too small, in order to allow sufficient sampling to reduce the noise when solving the synthetic equation.

To address this problem, we quickly test the intermittent GSIS in the linearized Poiseuille flow between two parallel plates, that is, we use the solutions of the synthetic equation to guide the evolution of traditional kinetic solver in every $N$ steps. Without losing of generality, the following steady-state BGK equation is used~\cite{LeiJCP2017}:
\begin{equation}\label{BGK_Poiseuille}
v_1\frac{\partial
	{h}}{\partial{x}}={\delta_{rp}}[2uv_3f_{eq}-h]+v_3f_{eq},
\end{equation}
where $f_{eq}(\bm{v})={\exp(-|\bm{v}|^2)}/{\pi^{3/2}}$ is the equilibrium velocity distribution function, with $\bm{v}=(v_1,v_2,v_3)$ being the three-dimensional molecular velocity; $h(x,\bm{v})$ is the perturbed velocity distribution function, and $x$ is the spatial coordinate perpendicular to the two parallel plates; the macroscopic flow velocity parallel to the two plates is $u=\int{v_3h}d\bm{v}$; the rarefaction parameter is $\delta_{rp}$ (proportional to the inverse Knudsen number). The synthetic equation is 
\begin{equation}\label{BGK_diffusion}
\frac{\partial^2 u}{\partial x^2}=
-\delta_{rp}
-\frac{\partial^2  }{\partial x^2} \int  h(x,v)(2v_1^2-1)v_3 d\bm{v}.
\end{equation}
The general procedure of GSIS and the conventional iterative scheme (CIS) for the kinetic equation is elaborated in Ref.~\cite{LeiJCP2017}. The BGK equation is solved by the second-order upwind finite difference scheme implicitly (\textbf{the effective time step is exactly the mean collision time, $1/\delta_{rp}$ which is roughly the inverse of Knudsen number Kn}), and the synthetic equation is also solved second-order upwind finite difference scheme.

\begin{figure}[t]
	\centering
	\includegraphics[width=0.49\textwidth]{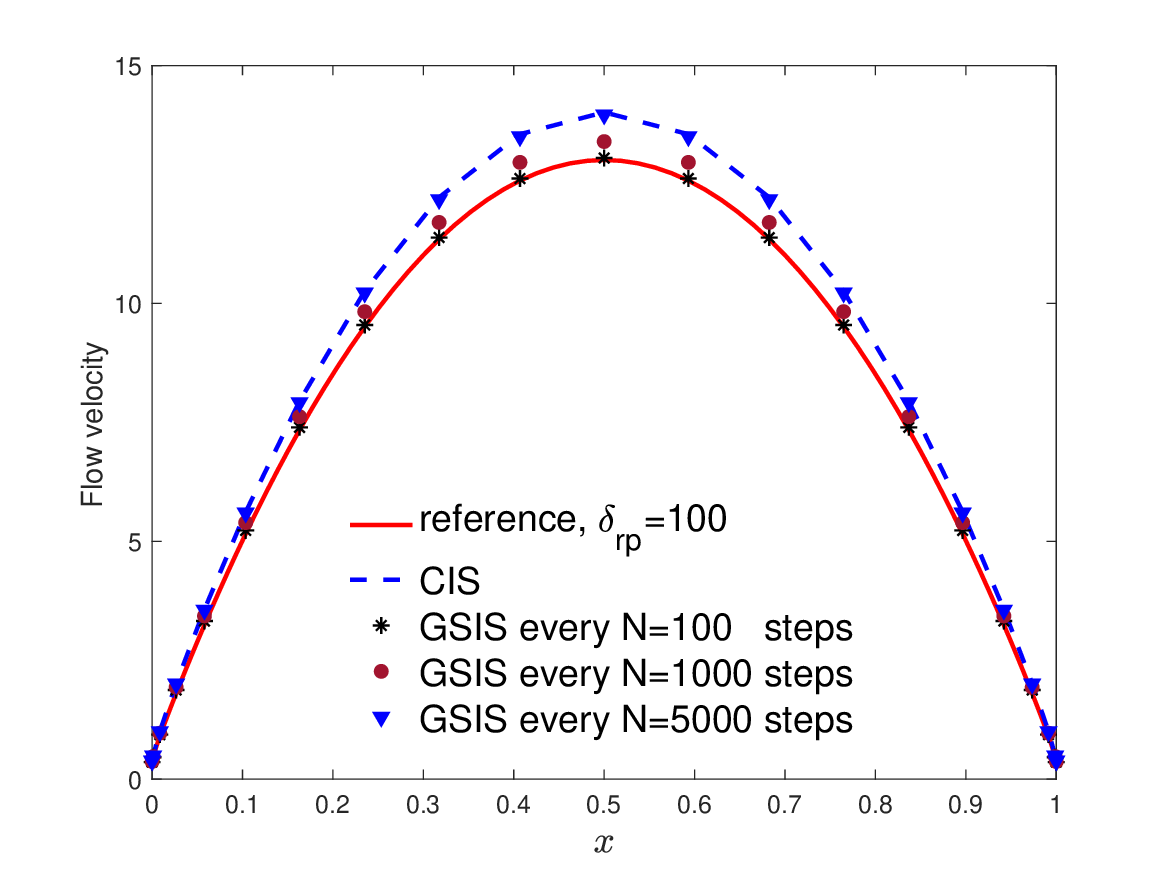}
	\includegraphics[width=0.49\textwidth]{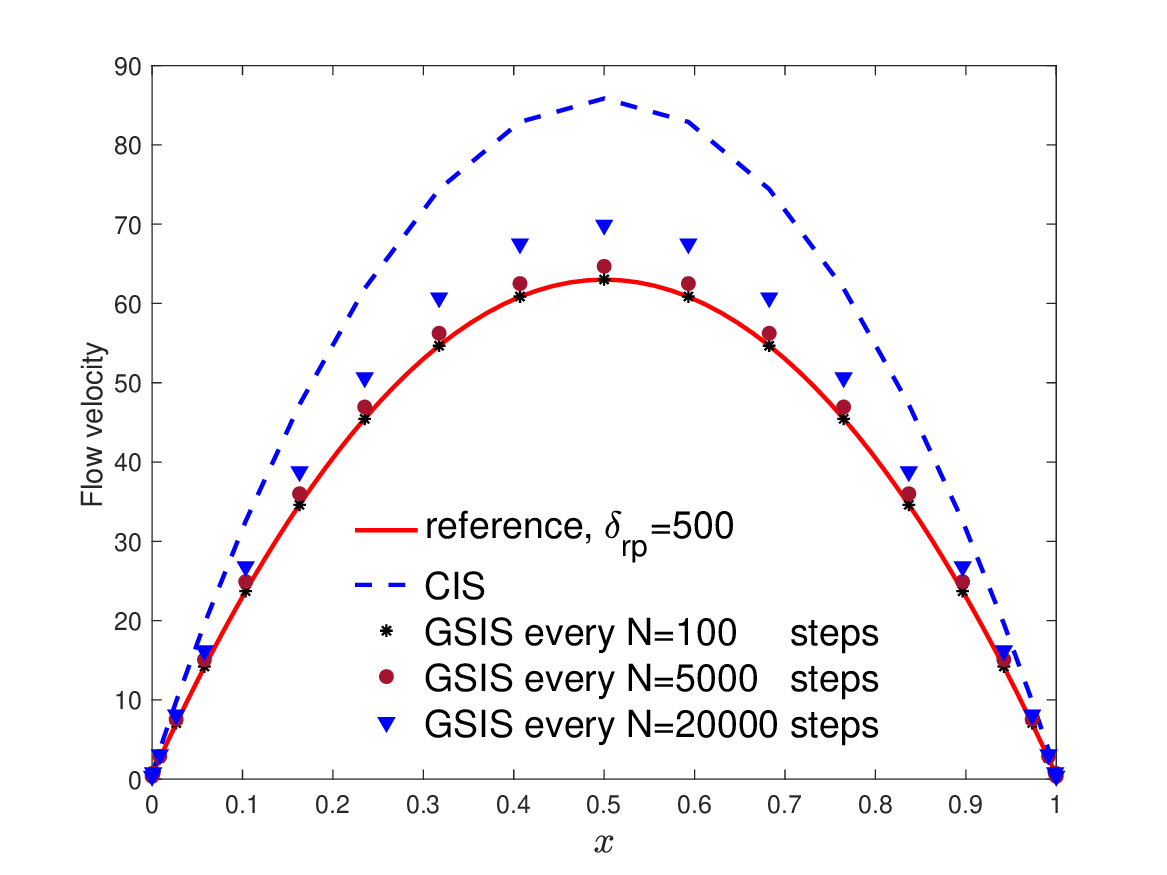}
	\caption{Numerical tests of the intermittent GSIS in the linearized Poiseuille flow, where the GSIS is applied to the CIS in every $N$ steps. Reference solutions are obtained from GSIS with refined spatial grids~\cite{LeiJCP2017}, while other solutions are obtained when the spatial domain $x\in[0,1]$ is discretized non-uniformly with 21 points. 
	}
	\label{Poiseuille_test}
\end{figure}

Figure~\ref{Poiseuille_test} shows the effects of intermittent GSIS, when the rarefaction parameter is small, and the flow is in the continuum regime. When $\delta_{rp}=100$,  the CIS with coarse spatial grid finds the wrong velocity profile after solving the BGK equation for 15,000 times (the convergence criterion is that the relative error in flow velocity between two consecutive iterations is less than $10^{-6}$). However, with the synthetic equation applied in every step, the original GSIS obtains the correct solution after solving the BGK equation for only 21 times, clearly demonstrating the fast convergence and asymptotic-preserving properties. If the GSIS is applied every $N=100$ steps a few times, the correct velocity profile can still be found. This is because the numerical error associated with the CIS does not have adequate time to develop. When $N=1000$, the numerical error in CIS gradually develops, and the final velocity profile obtained from the intermittent GSIS slightly deviates from the reference one. When $N=5000$, the numerical error in CIS has sufficient time to develop, and the intermittent GSIS is reduced to the pure CIS.  
When $\delta_{rp}=500$, the pure CIS introduces even larger numerical error, since the spatial cell size is much larger than the mean free path. Worse still, it needs 126,914 iteration steps, contrasting with the 18 steps demanded by the pure GSIS.
When applying the GSIS in every 100 (or 1,000, not shown for clarity) steps, the final intermittent GSIS solution agrees well with the reference one. However, as $N$ is increased to 5,000 and 20,000, the numerical error gradually develops in the intermittent GSIS. 
On the other hand, when $\delta_{rp}$ is small, e.g., the flow is in the  transition or free-molecular regime, the pure CIS is fast and accurate, and the use of intermittent GSIS will produce the correct solution, not matter how large the value of $N$ is (not shown in the figure for clarity).

From the Fourier stability analysis~\cite{Su2020SIAM}, we know that the iteration in pure CIS has a spectral radius of $1-\Kn^2/2$ when the Knudsen number $\Kn$ is small. This implies that after one iteration, the error is reduced by a mount of $\Kn^2$. Therefore, roughly speaking, the total iteration steps in pure CIS are proportional to $\delta_{rp}^2$ to find the steady-state solution (which, however, might subject to numerical errors if the spatial resolution is not high enough). This also means that, the iteration steps for the numerical error in the intermittent GSIS to emerge is roughly proportional to $\delta_{rp}^2$.
This is indeed supported by the numerical results in Fig.~\ref{Poiseuille_test}: when $\delta_{rp}=100$, the numerical error in the intermittent GSIS emerges when $N=1000$, while that of $\delta_{rp}=500$ in the intermittent GSIS is about 25 times larger, i.e., the error emerges when $N\approx20,000$. Therefore, the numerical dissipation in the pure CIS will be eliminated by the intermittent GSIS, e.g., if we choose $N\approx100$, even when the spatial cell size is about 100 times larger than the mean free path.

\section{The direct intermittent GSIS-DSMC coupling}

It can be concluded from the above preliminary tests that, applying the GSIS in about every 100 steps (when the effective time step is the mean collision time) will not only facilitate the asymptotic-preserving property of the kinetic solver, but also boost the convergence to the steady state, as compared to the original kinetic solver. The same should hold for the DSMC too. 

\begin{figure}[t]
	\centering
	\includegraphics[width=0.7\textwidth]{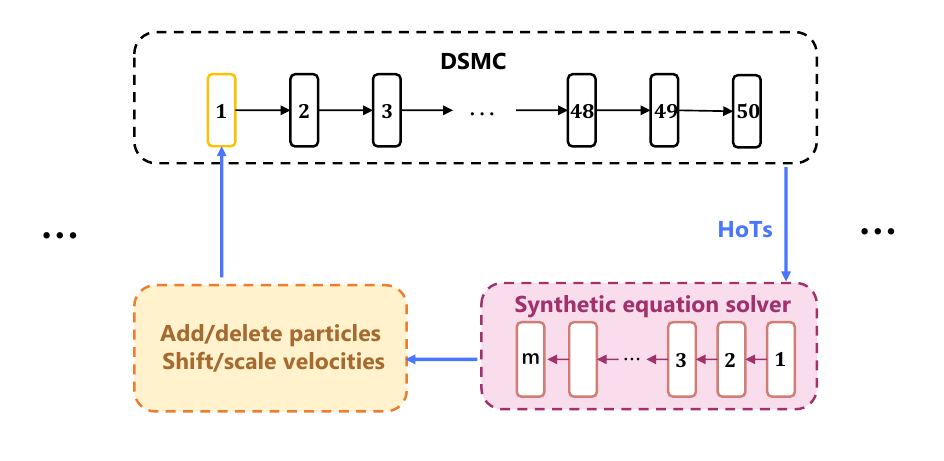}
	\caption{
 Flowchart of the direct intermittent GSIS-DSMC solver in a unit cycle, which is repeated until convergence.
	}
	\label{flowchart}
\end{figure}

Thus, on top of our recent work~\cite{Luo2024arXiv},  we design the direct intermittent GSIS-DSMC solver (DIG) in Fig.~\ref{flowchart}. The `direct' means that the DIG is based on the standard DSMC, replacing the modification of the particle distribution via the complicated Wild sum in time-relaxed Monte Carlo method~\cite{Pareschi2001}. Main steps of the DIG solver are given below:
\begin{enumerate}

	\item Solve the traditional Navier-Stokes equation to get initial flow field, and initialize the DSMC simulation to the Maxwellian distribution, with the obtained density, velocity, temperature. 

    \item Run the standard DSMC for 50 steps, and get the time-averaged macroscopic quantities. To reduce the thermal fluctuation, the exponentially weighted moving time averaging method with 100 samples~\cite{Jenny2010JCP} is employed, see Appendix~\ref{appendix}.
    
    \item Solve the macroscopic synthetic equation~\eqref{eq:Navior-Stokes} for $m=500\sim2000$ steps, or till the relative error in conservative variables between two consecutive steps smaller than $10^{-5}$. The boundary condition and numerical method for the synthetic equation are detailed in Ref.~\cite{Luo2024arXiv}. 

	\item Correct the particle distribution in DSMC to reflect the change of density, velocity, and temperature as specified by the solutions of the synthetic equation, see the details in section 3.2 in Ref.~\cite{Luo2024arXiv}. 
	
	\item Repeat steps 2-4 (i.e., Fig.~\ref{flowchart}) until convergence.
\end{enumerate}

\subsection{Hypersonic flows passing over a cylinder}

Consider the argon gas flow of Mach number $\Ma=5$ passing over a cylinder. 
The Knudsen number is defined as
\begin{equation}
\text{Kn} =\frac{\lambda}{L} =\frac{\mu_{0}(T_0)}{p_{0} L}\sqrt{\frac{\pi k_B T_{0}}{2m}},
\end{equation}
where the molecular mean free path is $\lambda$, the characteristic flow length is the cylinder diameter $L$, the inflow pressure and temperature are respectively $p_0$ and $T_0$. The variable hard-sphere model is used in DSMC, where the viscosity is calculated as $\mu(T)=\mu_{0}(T_0)\times(T/T_{0})^\omega$ with the exponent $\omega = 0.81$. The Maxwellian diffuse boundary condition is applied at the cylinder surface with the temperature $T_0$. 

\begin{figure}[p]
    \centering
    \includegraphics[width=0.49\textwidth]{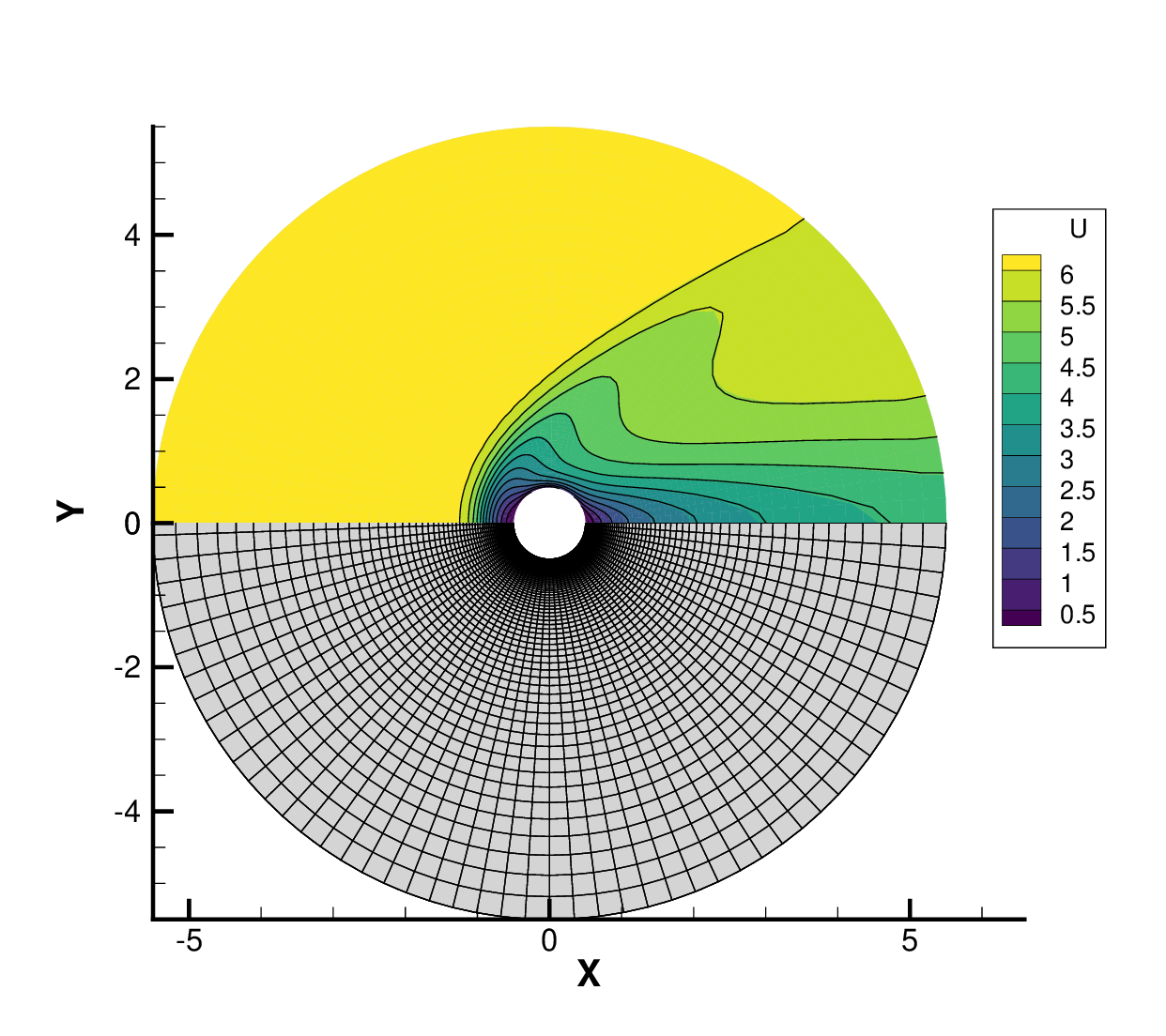}
    \includegraphics[width=0.49\textwidth]{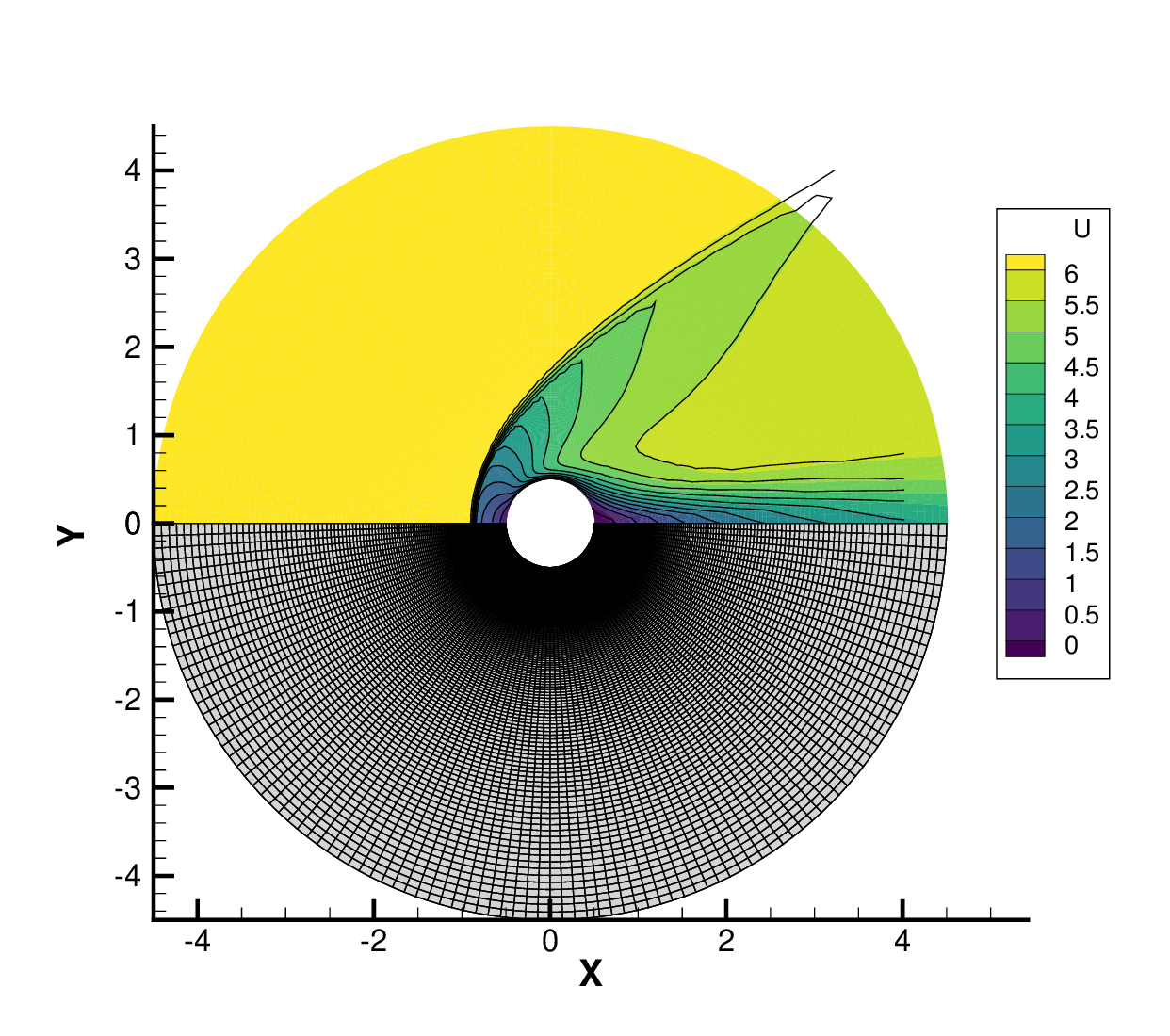}\\
        \includegraphics[width=0.49\textwidth]{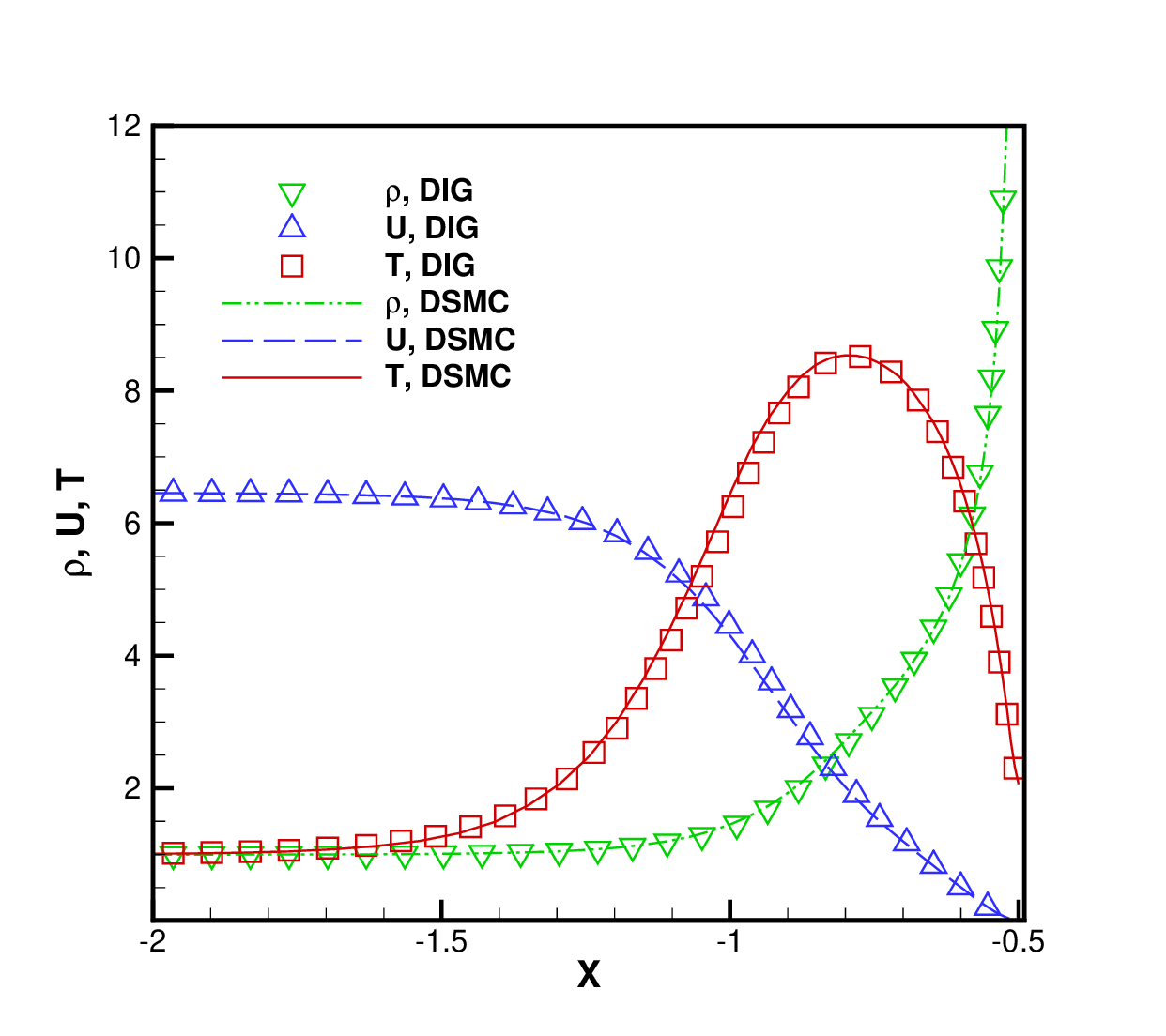}
    \includegraphics[width=0.49\textwidth]{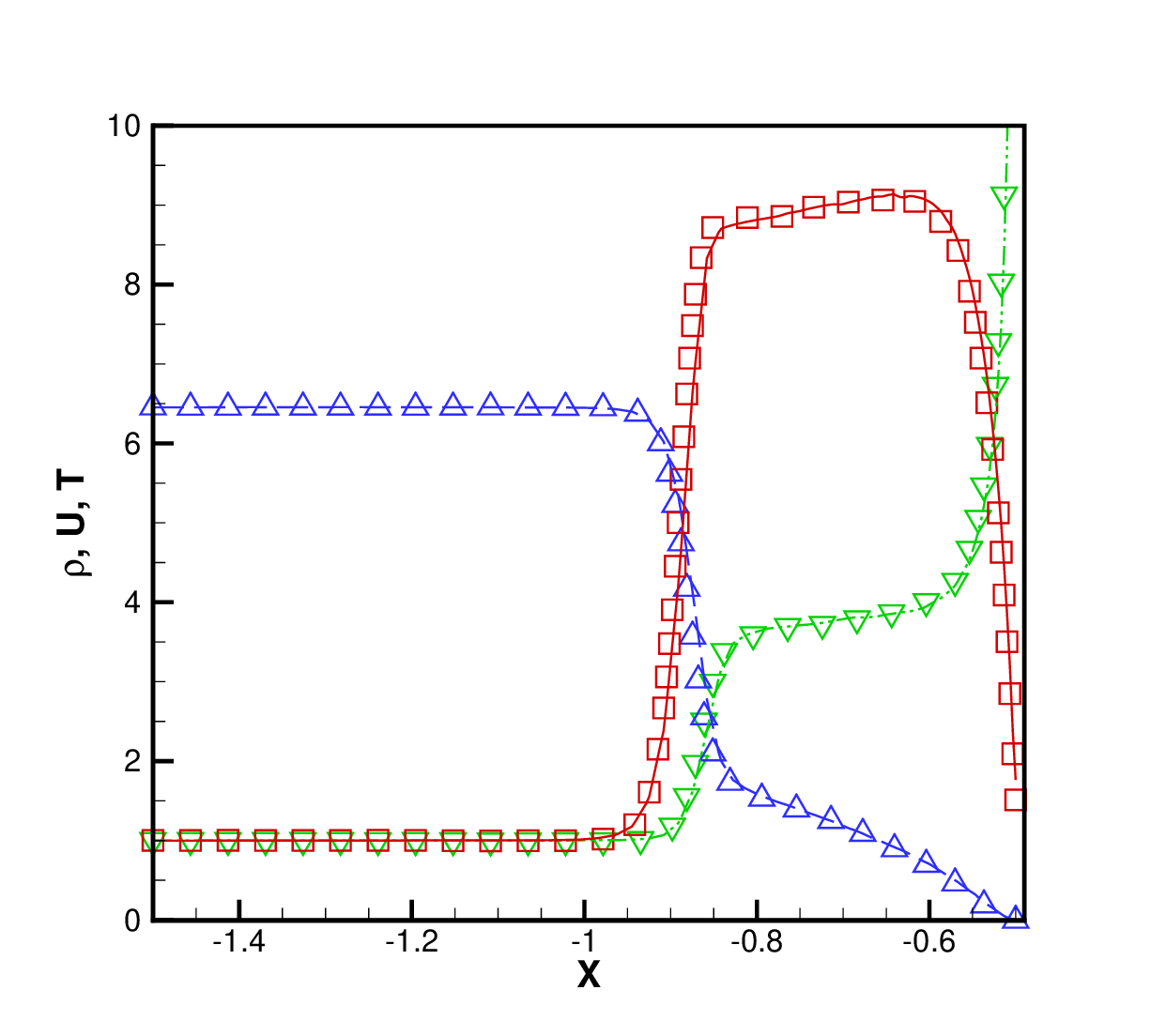}\\
    \includegraphics[width=0.49\textwidth]{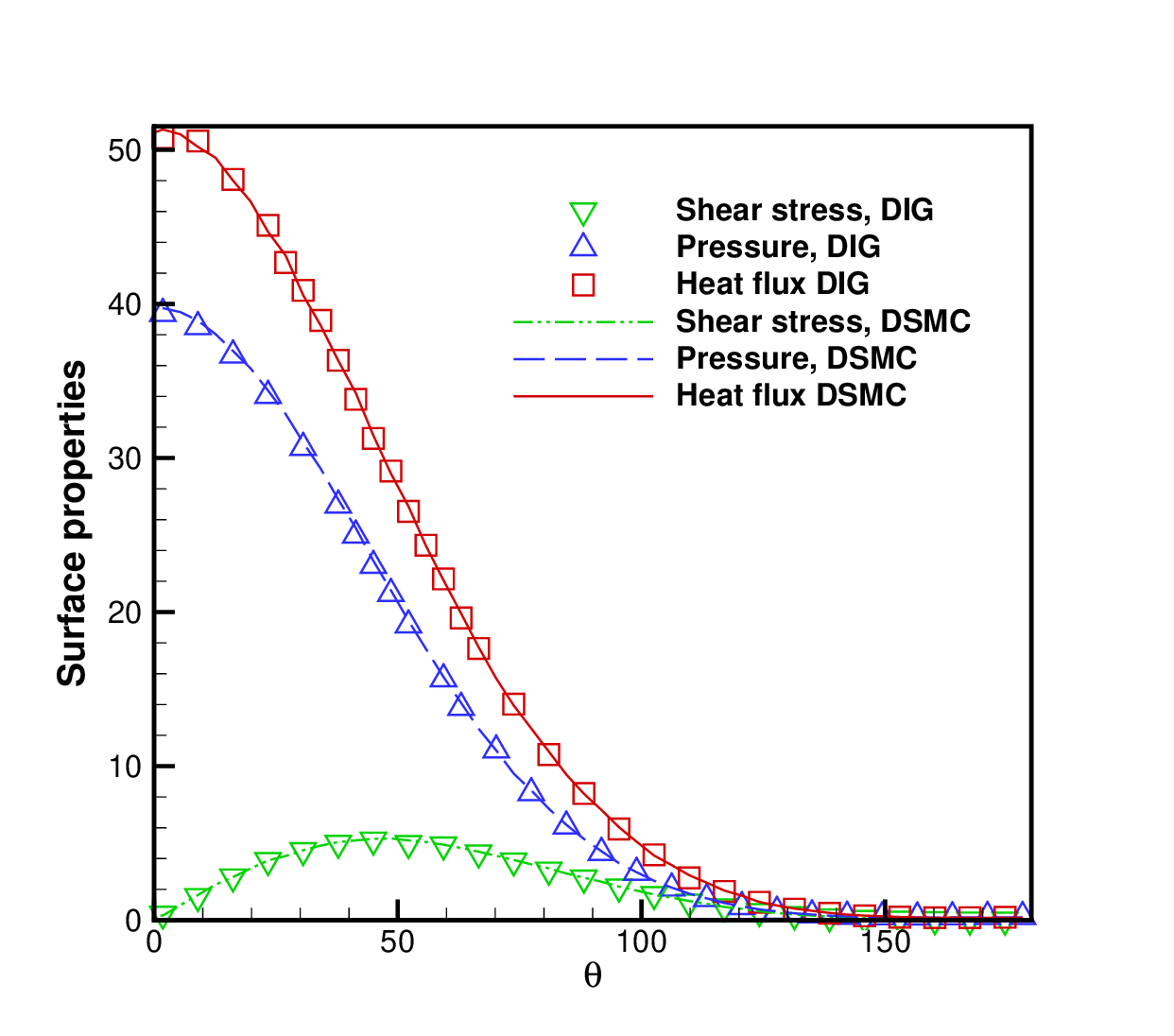}
    \includegraphics[width=0.49\textwidth]{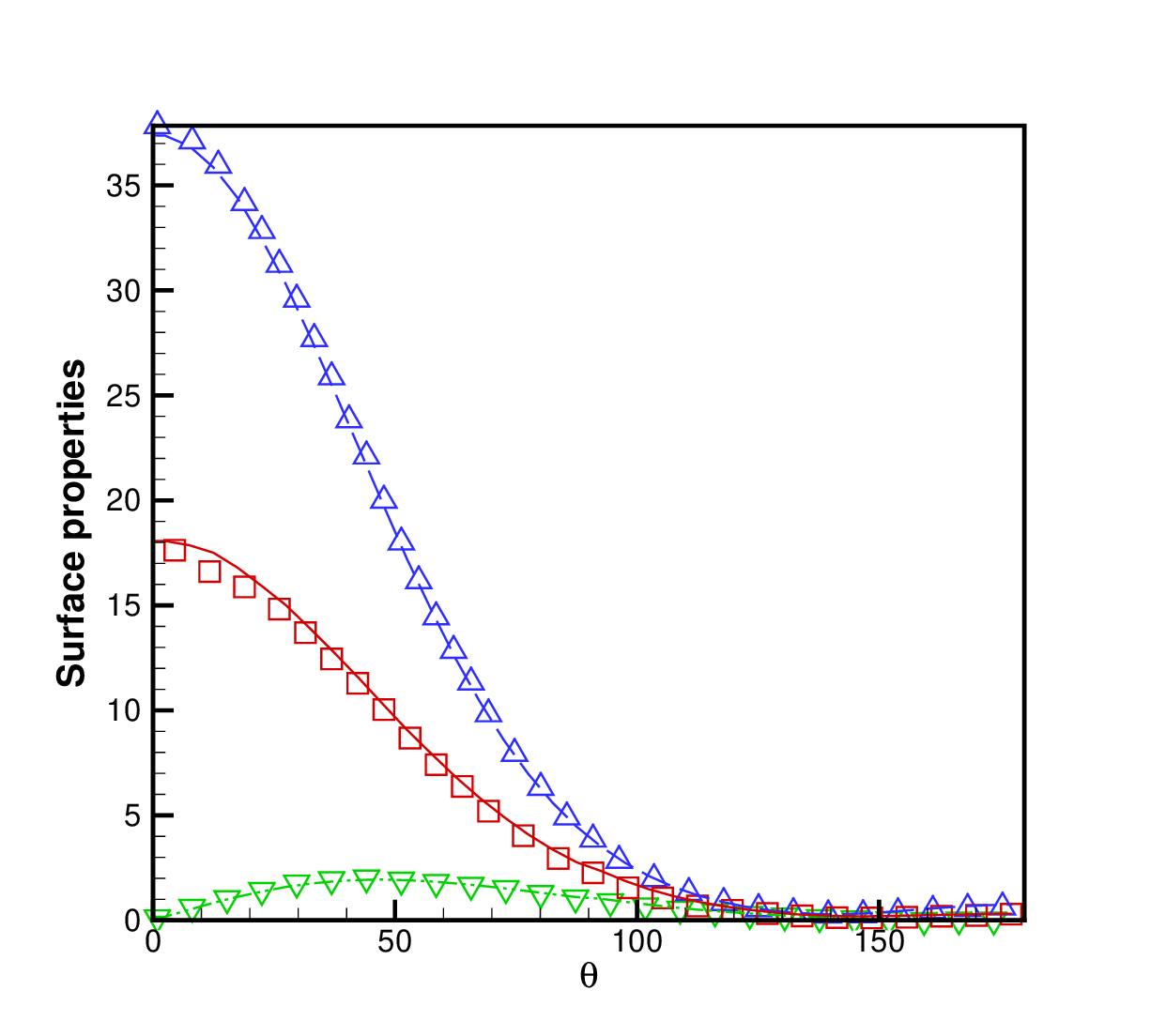}
    \caption{
    The hypersonic flow passing over a cylinder, when $\Kn=0.1$ (left column) and $\Kn=0.01$ (right column). (First row) The mesh and velocity contour. The results of DIG and DSMC are shown as background and lines, respectively. (Second row) Density, horizontal velocity, and temperature along the stagnation stream line in windward side of the cylinder. (Third row) Shear stress, pressure, and heat flux on the cylinder surface, where the angle $\theta$ ($^\circ$) is measured from the leading edge of the cylinder. 
}
    \label{fig:contourmesh}
\end{figure}

The computational domain is an annulus with an inner circle being the cylinder surface and outer circle being the far field. 
The radius of the outer circle is $5.5L$ for $\Kn=0.1$ and $4.5L$ for $\Kn=0.01$, and that of the inner circle is $0.5L$. The non-uniform structured mesh is used, and the total cell numbers in the circumferential and radial directions are $M$ and $N$, respectively. As shown in Table~\ref{tab1} and Fig.~\ref{fig:contourmesh}, when $\Kn=0.1$,  $M=100$ and $N=64$. When $\Kn=0.01$, the physical grid is set as $M=200$, $N=200$. The height of the first layer grid is $\Delta h = 0.2\lambda$; such a small first layer height is necessary to capture the surface heat flux. In all cases, 100 particles are assigned in each cell initially.
Furthermore, when $\Kn=0.1$, a CFL number of 0.2 is employed in the pure DSMC and DIG, and is increased to 0.5 for DIG when $\Kn=0.01$. The CFL number in the macroscopic solver is 5. 

\begin{table}[t]
 \centering
 \caption{\label{tab1}The computational overhead in the hypersonic flow around a cylinder. The computational time is expressed in core$\times$hours. Results with * were obtained from the SPARTA program, using non-uniform Cartesian grids and uniform initial conditions. In contrast, other simulations were initialized using solutions from the traditional Navier-Stokes equations.}
\begin{tabular}{c c c c c c c c}\toprule
 \multirow{2}{*}{Kn}  & \multirow{2}{*}{Methods} & \multirow{2}{*}{CFL} & \multirow{2}{*}{$N_{\text{cell}}$}  & \multicolumn{2}{c}{Transition state}  &  \multicolumn{2}{c}{Steady state} \\ \cmidrule(r){5-6} \cmidrule(r){7-8}
 ~ & & ~ & ~  & steps&time&steps   &    time\\ \hline
 \multirow{2}{*}{0.1} & DSMC & \multirow{2}{*}{0.2} & \multirow{2}{*}{$100\times 64$}  &    700   & 0.3 &  10000 &     4.7     \\
~  & DIG &    & ~   &  400   & 0.3 &  5000 &      3.2     \\  \addlinespace
 \multirow{2}{*}{0.01}  & DSMC$^*$ & 0.2&  2,010,616  &    50000   & 300 &  50000 &  295     \\
~   & DIG  & 0.5 & $200\times 200$ &    400   & 2.1 &  3000 &      15.2     \\  \addlinespace
\bottomrule
\end{tabular}
\end{table}

\begin{figure}[t]
    \centering
    \includegraphics[width=0.49\textwidth]{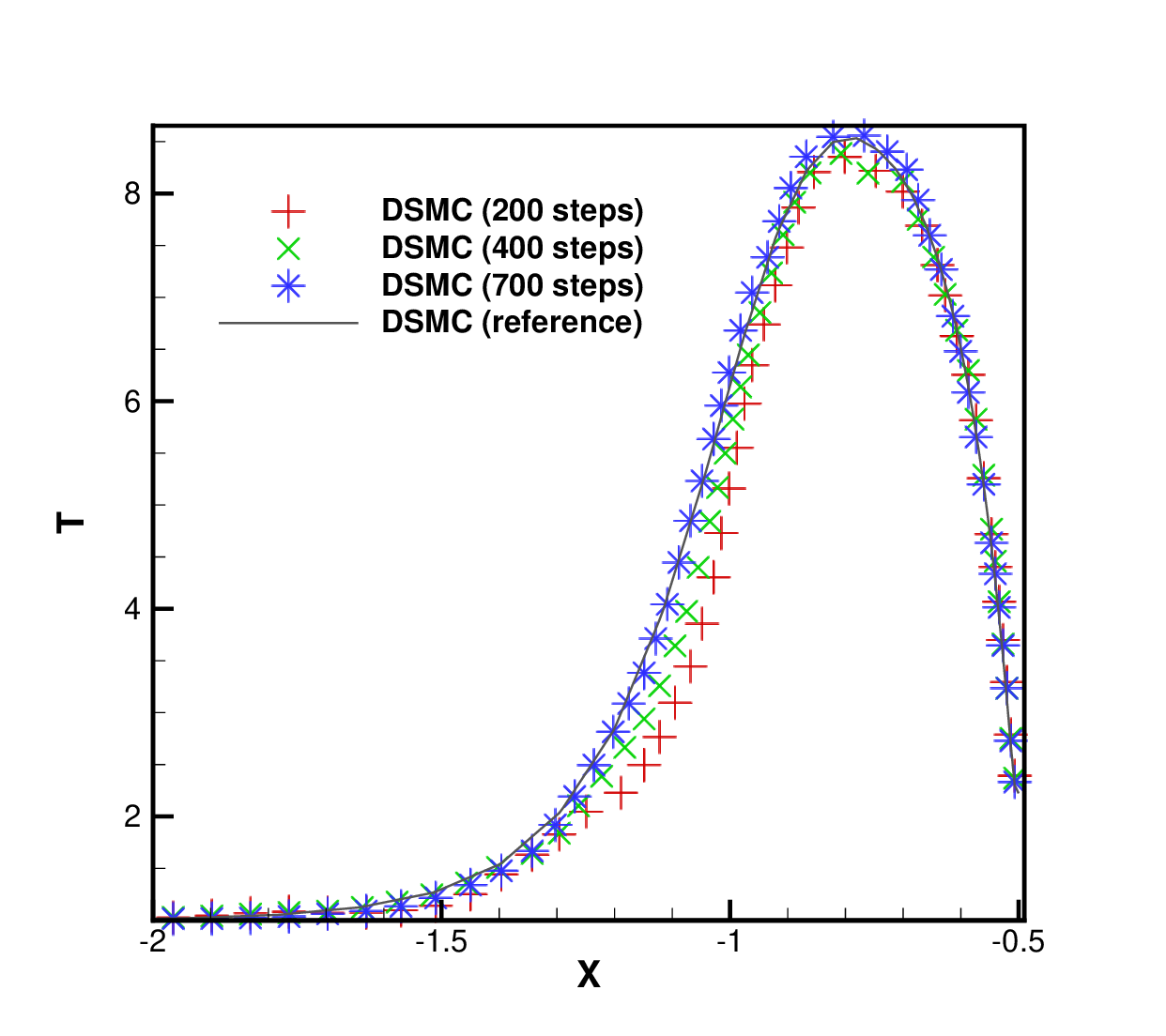}
    \includegraphics[width=0.49\textwidth]{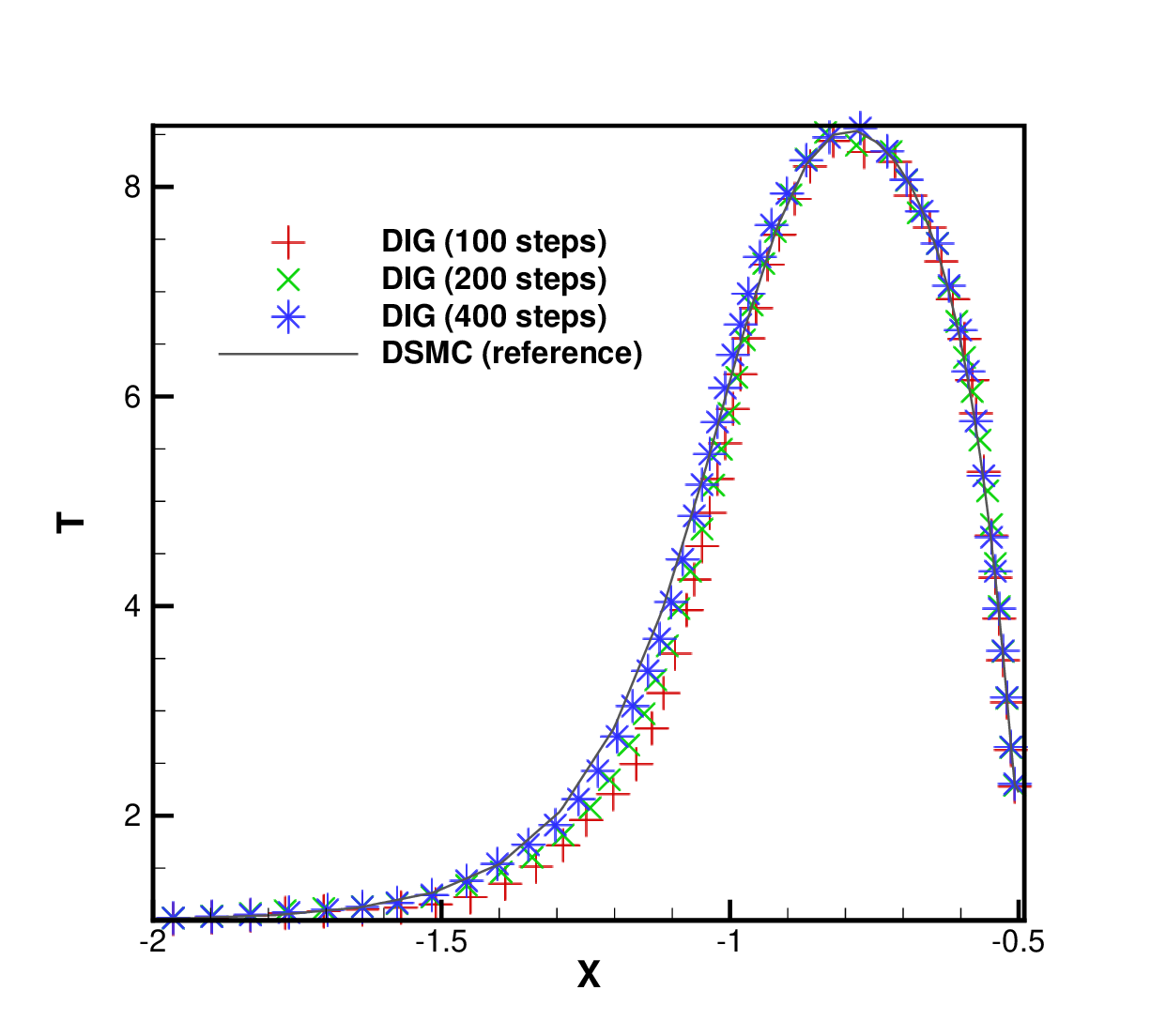}
    \includegraphics[width=0.49\textwidth]{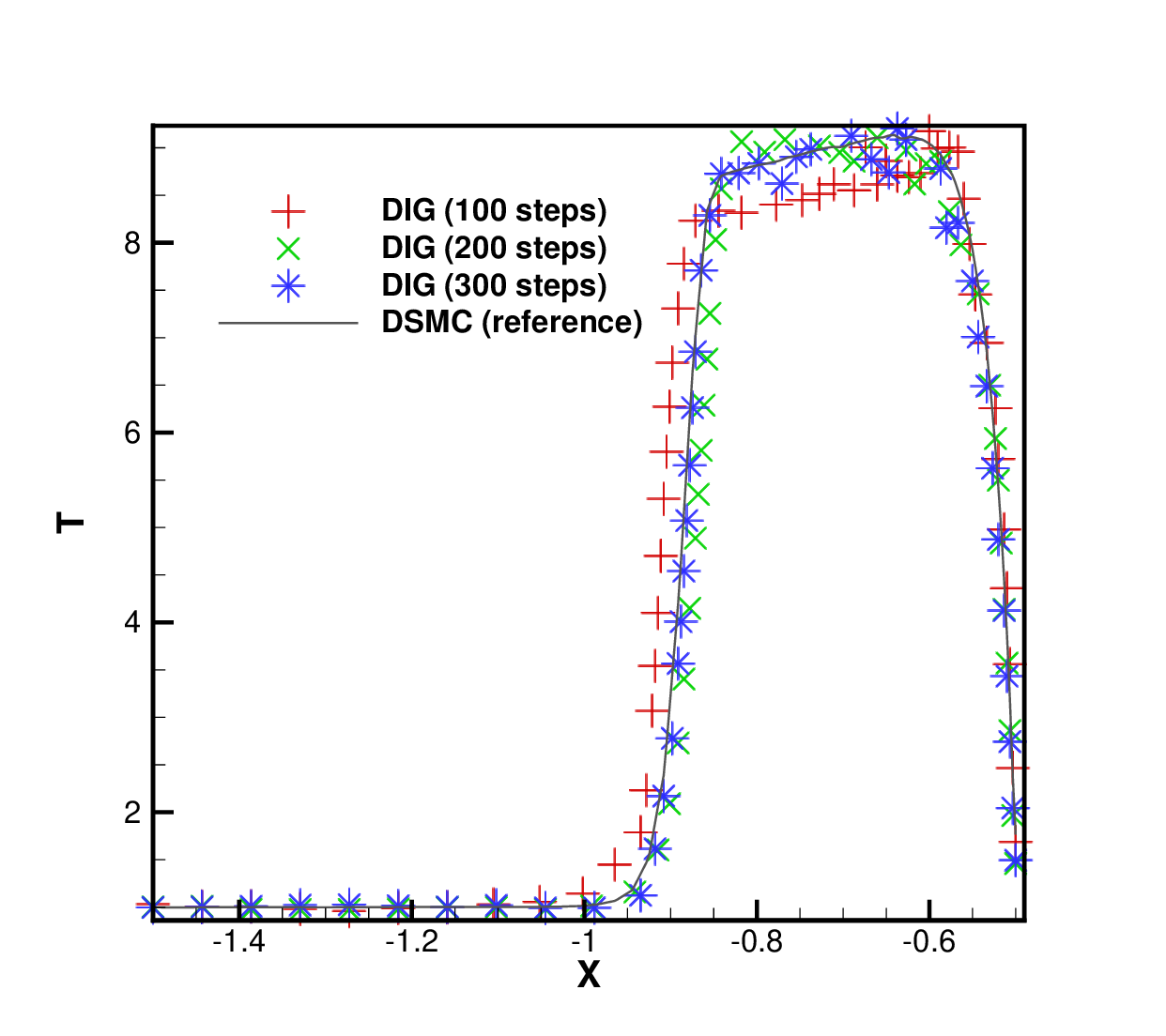}
    \caption{
    The evolution of the temperature in horizontal direction in the windward side of the cylinder surface when $\Kn=0.1$ (first row) and $\Kn=0.01$ (second row). 
    }
    \label{fig:evolution_cylinder}
\end{figure}

Given the GSIS's characteristic of fast convergence, the transition state within the DIG is relatively brief. Figure~\ref{fig:evolution_cylinder} illustrates the evolution of the temperature in the horizontal direction along the windward side of the cylinder surface when $\Kn=0.1$ and $\Kn=0.01$. When $\Kn=0.1$, DSMC necessitates 700 steps to reach the steady state, while DIG requires approximately 300 steps. However, when the Knudsen number decreases to 0.01, while the DSMC needs 50,000 steps to reach the steady state, the DIG requires only 300 steps.

Figure~\ref{fig:contourmesh} shows the steady-state velocity contours, the density, temperature, and velocity along the stagnation stream line, and the flow variables along the cylinder surface. It can be seen that the DIG results agree well with the DSMC, especially when $\Kn=0.01$ where the spatial resolution in DIG is much smaller than DSMC. This is attributed to the GSIS's asymptotic-preserving property, where accurate solutions can be found even using coarse spatial grids.

Thanks to the fast convergence and asymptotic-preserving properties brought by the macroscopic synthetic equation, Table~\ref{tab1} confirms that the DIG is much more efficient than the DSMC in the near-continuum flow regime. That is, when $\Kn=0.01$, since the iteration step is reduced by several orders of magnitude, the time for solving the synthetic equation can be neglected. Consequently, the overall CPU time of DIG is smaller than DSMC by two orders of magnitude. However, in the transition flow regime, e.g., $\Kn=0.1$, although the number of iterations required in the transition state is reduced, the overall CPU cost of DIG is similar to the DSMC, due to the extra cost in solving the synthetic equation. 
Additionally, since the complicated time-relaxed Monte Carlo is replaced by the standard DSMC, the simulation time is reduced by several times when compared to our previous version of GSIS and DSMC coupling~\cite{Luo2024arXiv}.

\subsection{Lid-driven cavity flow}

We then test the DIG in the low-speed lid-driven cavity flow. 
The computational domain is a $L\times L$ square cavity. All solid walls have the same temperature $T_0$. The top lid of the cavity moves horizontally at a speed of $U_w =\sqrt{2}c_0$ when $\Kn\ge 0.01$. And in the near-continuum regime, to avoid turbulence, $U_w$ is reduced to 0.21$c_0$ and 0.42$c_0$, corresponding to $\Kn=2.63\times10^{-3}$ and $5.26\times10^{-4}$, respectively. 
As summarized in Table~\ref{tab2}, when $\Kn=0.1$, $50\times50$ uniform spatial grids are employed in both DIG and DSMC. 
When $\Kn=0.01$, $100\times100$ non-uniform grids are applied in DIG, which are refined near the solid walls, e.g., the first layer cell has a thickness of $\Delta {h}=0.002$, contrasting with the $500\times500$ uniform grids in DSMC. For cases with $\text{Re} = 100 $ and $\text{Re} = 1000 $, $150\times150$ non-uniform grids with the first layer thickness $\Delta h=0.001$ are employed in DIG.
At the beginning of simulation, each computational cell is populated with 100 simulation particles, and the velocities of all particles satisfy the Maxwell distribution function of $\rho=T=1,\bm{u}=0$. The CFL numbers of DIG and DSMC are also shown in Table~\ref{tab2}.

\begin{table}[t]
 \centering
\caption{\label{tab2}
Computational overhead in the lid-driven flow. The computational time is given in core hours.
``-'' means that the computational cost of DSMC is huge. 
}
  \begin{tabular}{c c c c c c c c c c}\toprule
 \multirow{2}{*}{Kn} & \multirow{2}{*}{Re}   &  \multirow{2}{*}{$U_w$} & \multirow{2}{*}{Methods} & \multirow{2}{*}{CFL}   & \multirow{2}{*}{$N_{\text{cell}}$}  & \multicolumn{2}{c}{Transition state}  &  \multicolumn{2}{c}{Steady state} \\ \cmidrule(r){7-8} \cmidrule(r){9-10}
 ~ & ~ & ~ & ~ &~       &~                          & steps&time&steps   &    time\\ \hline
 \multirow{2}{*}{0.1} & \multirow{2}{*}{17.73} & \multirow{2}{*}{1.41}&DSMC &   \multirow{2}{*}{0.2}   &  \multirow{2}{*}{$50\times50$}  &    800   & 0.027 & 1E4 &      0.32     \\
~   & ~ &~   & DIG     & ~            &    ~ &    300   & 0.029 &  1E4 &      0.38     \\  \addlinespace
 \multirow{2}{*}{0.01}& \multirow{2}{*}{177.3} & \multirow{2}{*}{1.41}&DSMC &  0.2   &  $500\times500$&  2E4   & 17    &  1E5 &      80        \\
~   & ~ &~    & DIG     & 0.2            & $100\times100$ &    600   & 0.24 &5E4&      18     \\  \addlinespace 
 \multirow{2}{*}{2.63E{-3}}& \multirow{2}{*}{100}&  \multirow{2}{*}{0.21}&DSMC  &-& -  &  -   & -    &  -&      -        \\
~   & ~&~    & DIG    & 0.5            &  $150\times150$ &   1000   & 1.1 & 5E4 &      48     \\  \addlinespace
 \multirow{2}{*}{5.26E{-4}}& \multirow{2}{*}{1000}&  \multirow{2}{*}{0.42}&DSMC  &-& -  &  -   & -    &  -&      -        \\
~   & ~&~    & DIG     & 0.5            &  $150\times150$ &   2000   & 2.9 & 5E4 &  71  
    \\  
\bottomrule
\end{tabular}
\end{table}

\begin{figure}[t]
    \centering
    \includegraphics[width=0.49\textwidth]{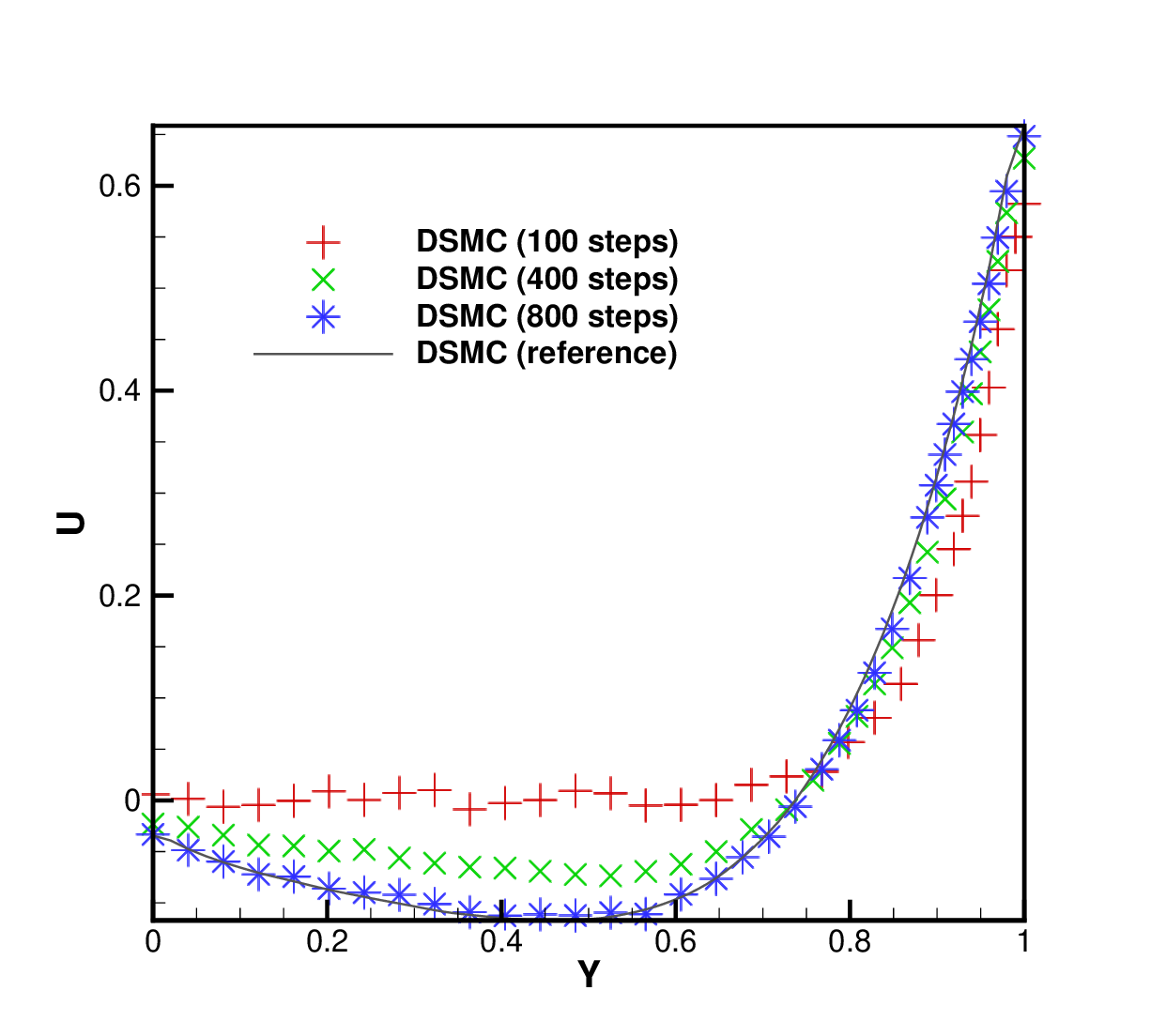}
    \includegraphics[width=0.49\textwidth]{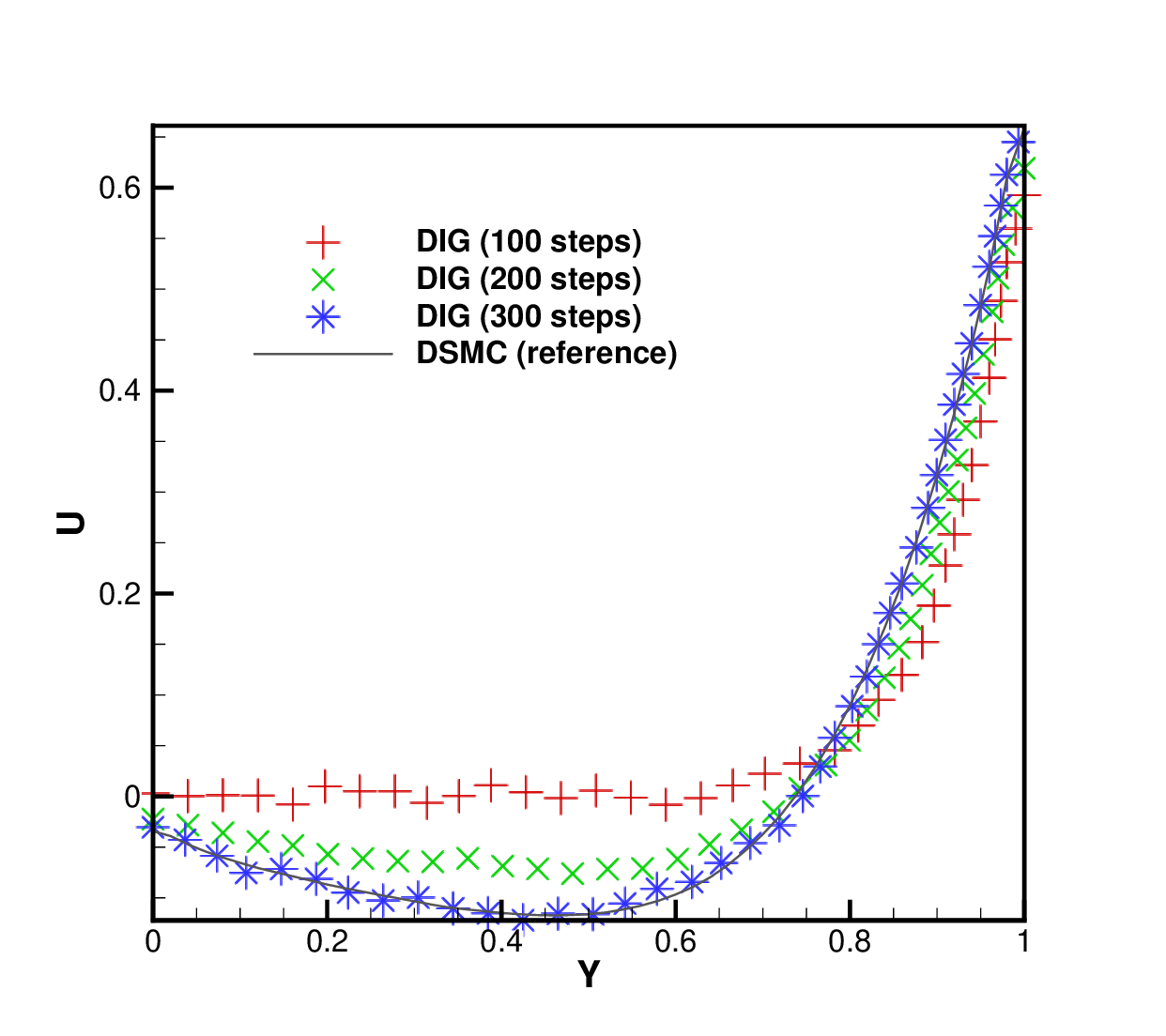}\\
    \includegraphics[width=0.49\textwidth]{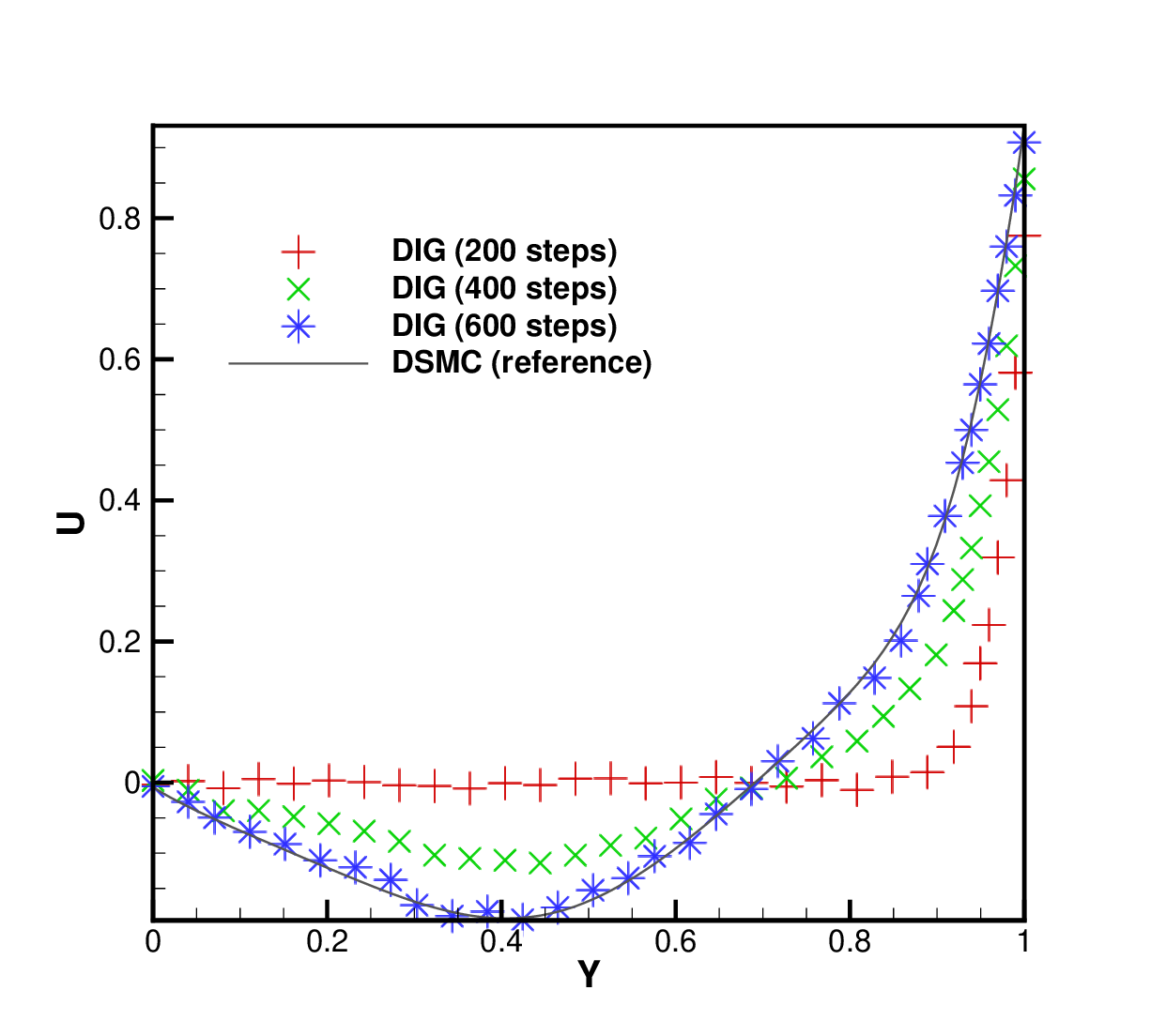}
    \caption{The evolution of horizontal velocities (at $x=0.5L$) in the transition stage in the lid-driven cavity flow, when $\text{Kn}=0.1$ (top) and 0.01 (bottom). }
    \label{fig:cavityevolution}
\end{figure}

The first row of Fig.~\ref{fig:cavityevolution} compares the evolution of horizontal velocity in the transition state, between the DSMC and DIG, when $\Kn=0.1$. Since the Knudsen number is not small, after a few hundreds of evolution steps, the steady states are reached in both schemes. Of course, DIG evolves slightly faster than the DSMC, due to the help of macroscopic synthetic equation. However, when the Knudsen number is small, e.g. $\Kn=0.01$ in the second row of Fig.~\ref{fig:cavityevolution}, DIG still needs a few hundreds of evolution steps to reach the steady state, but that of DSMC is much longer (not shown).

\begin{figure}[p]
    \centering
    \includegraphics[width=0.49\textwidth]{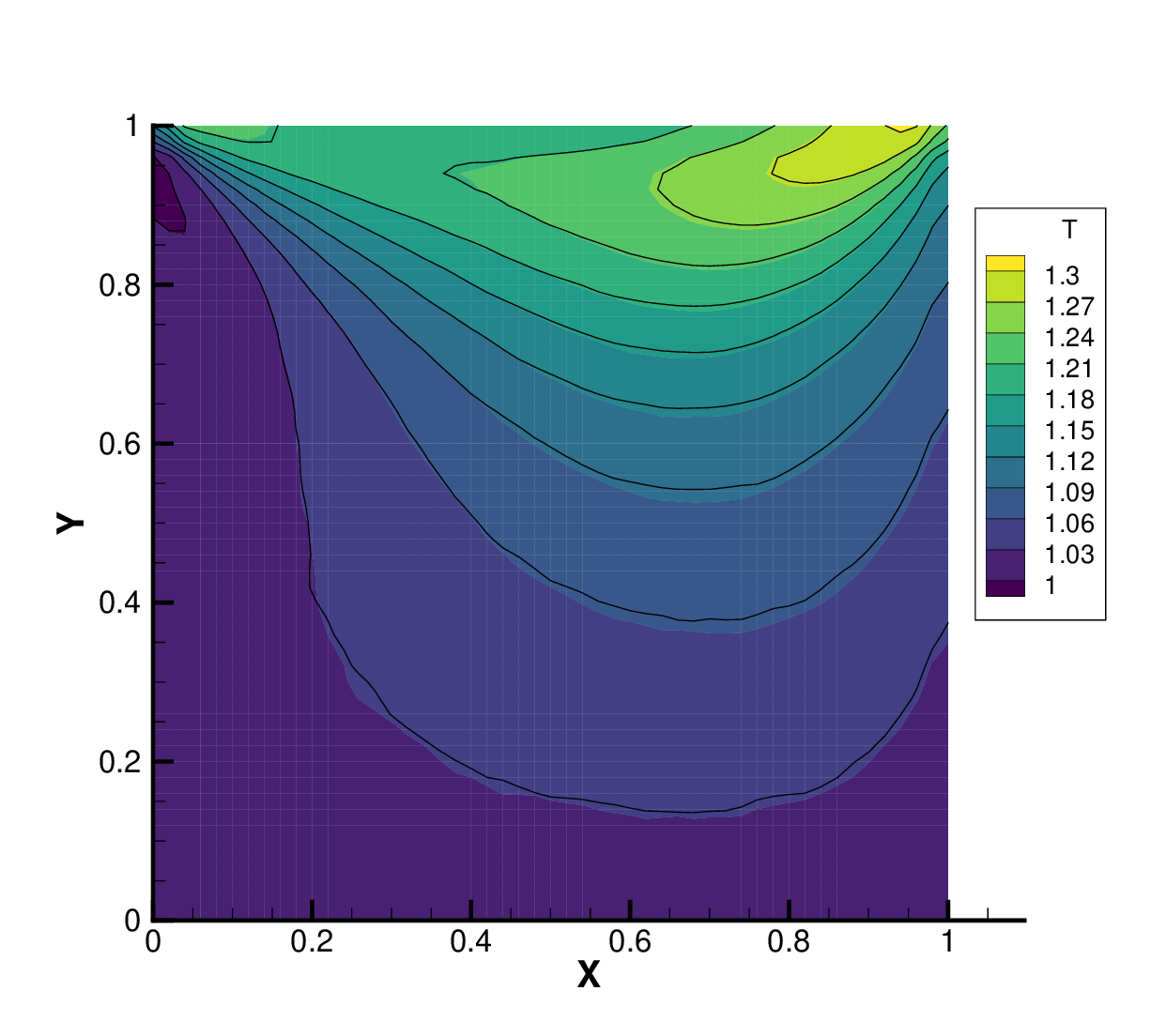}
    \includegraphics[width=0.49\textwidth]{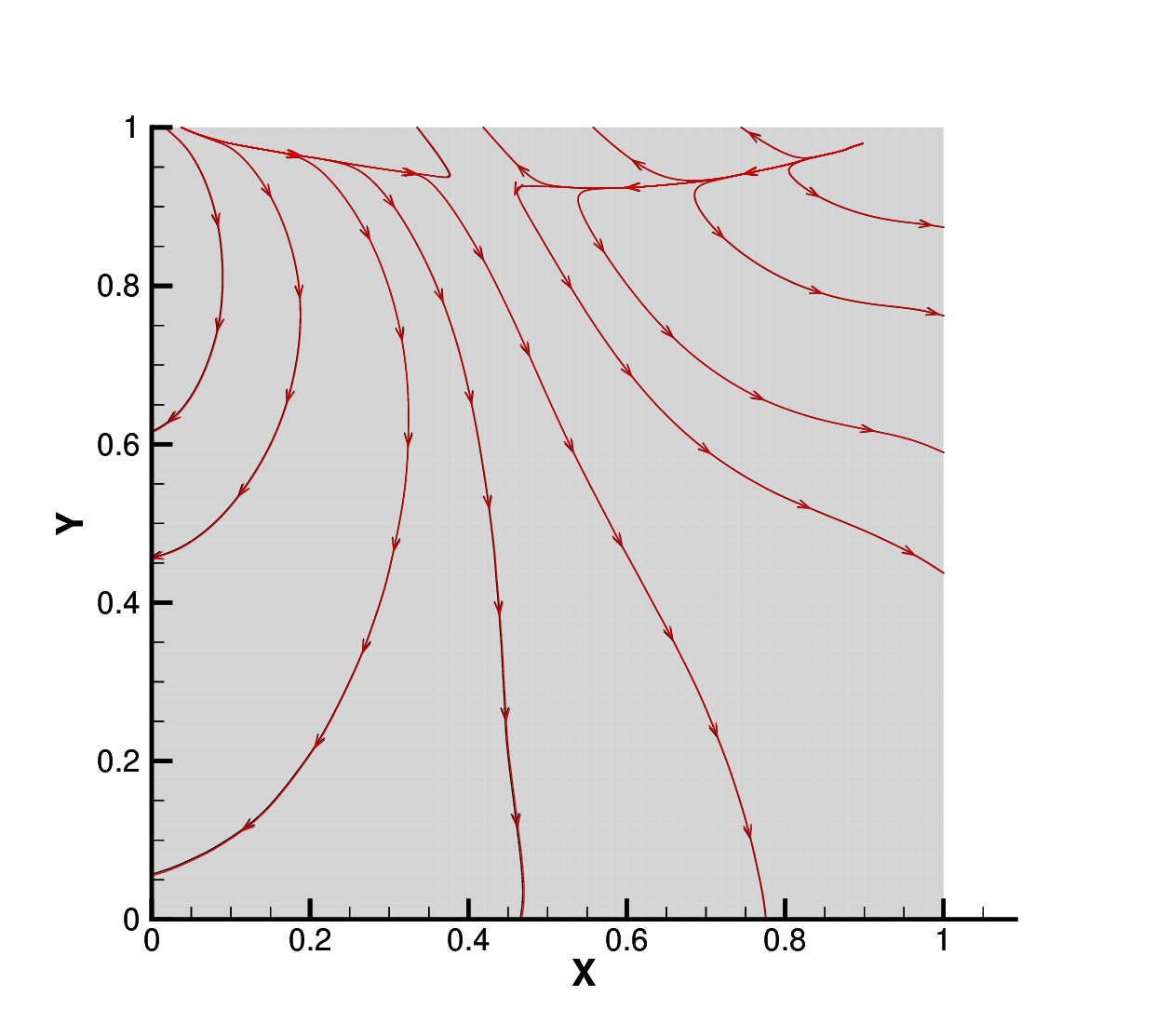}\\
    \includegraphics[width=0.49\textwidth]{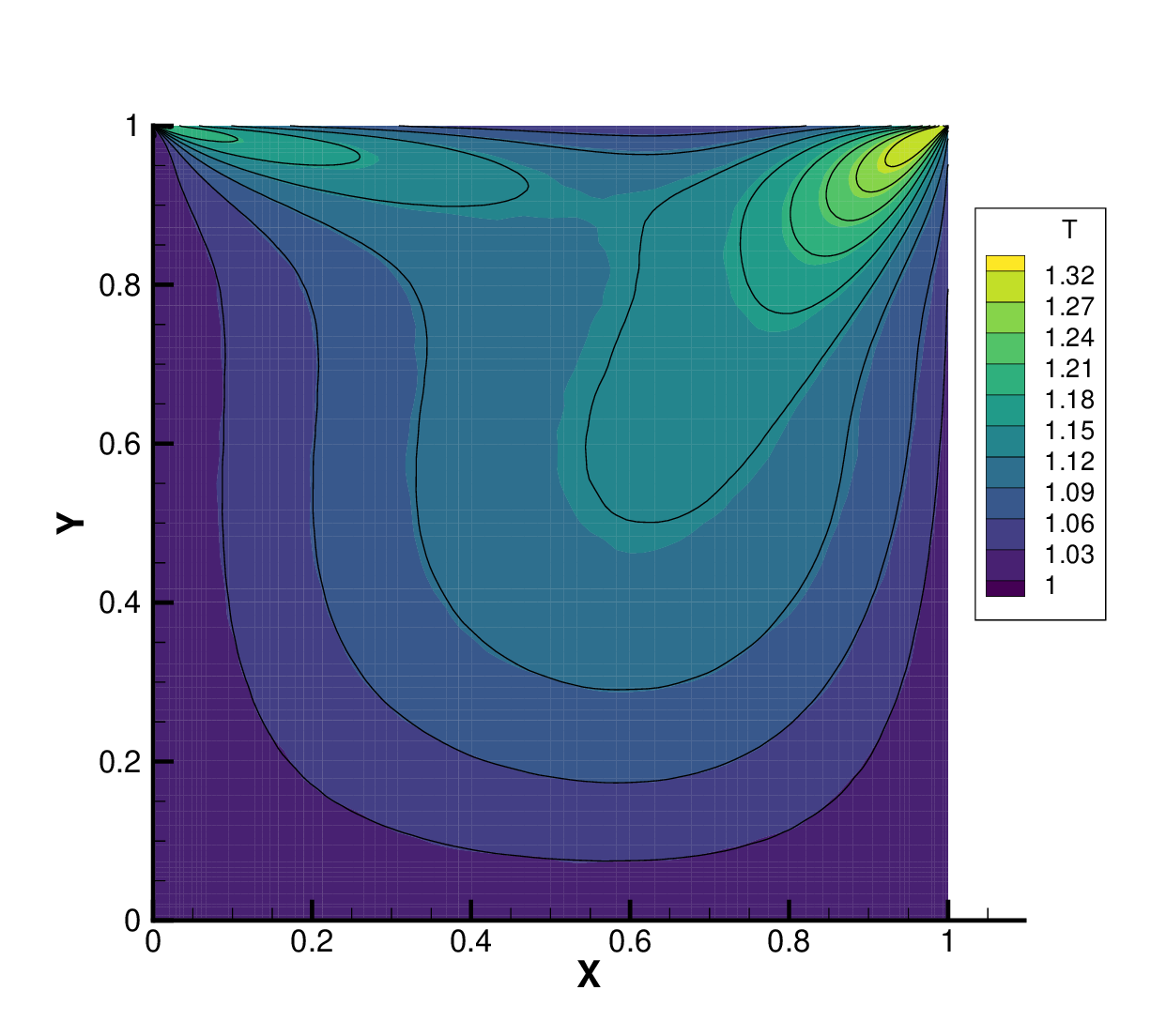}
    \includegraphics[width=0.49\textwidth]{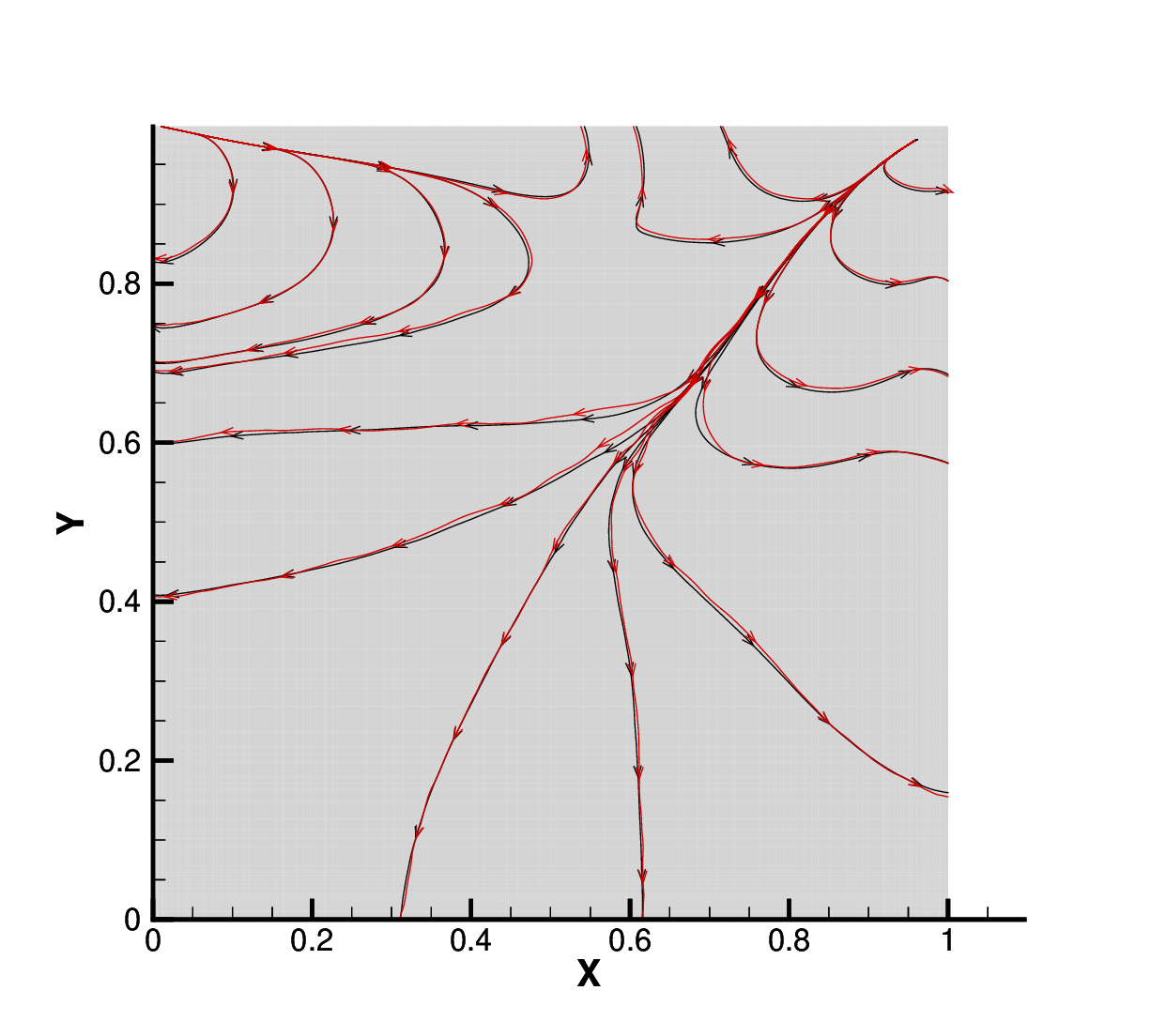}\\
      \includegraphics[width=0.49\textwidth]{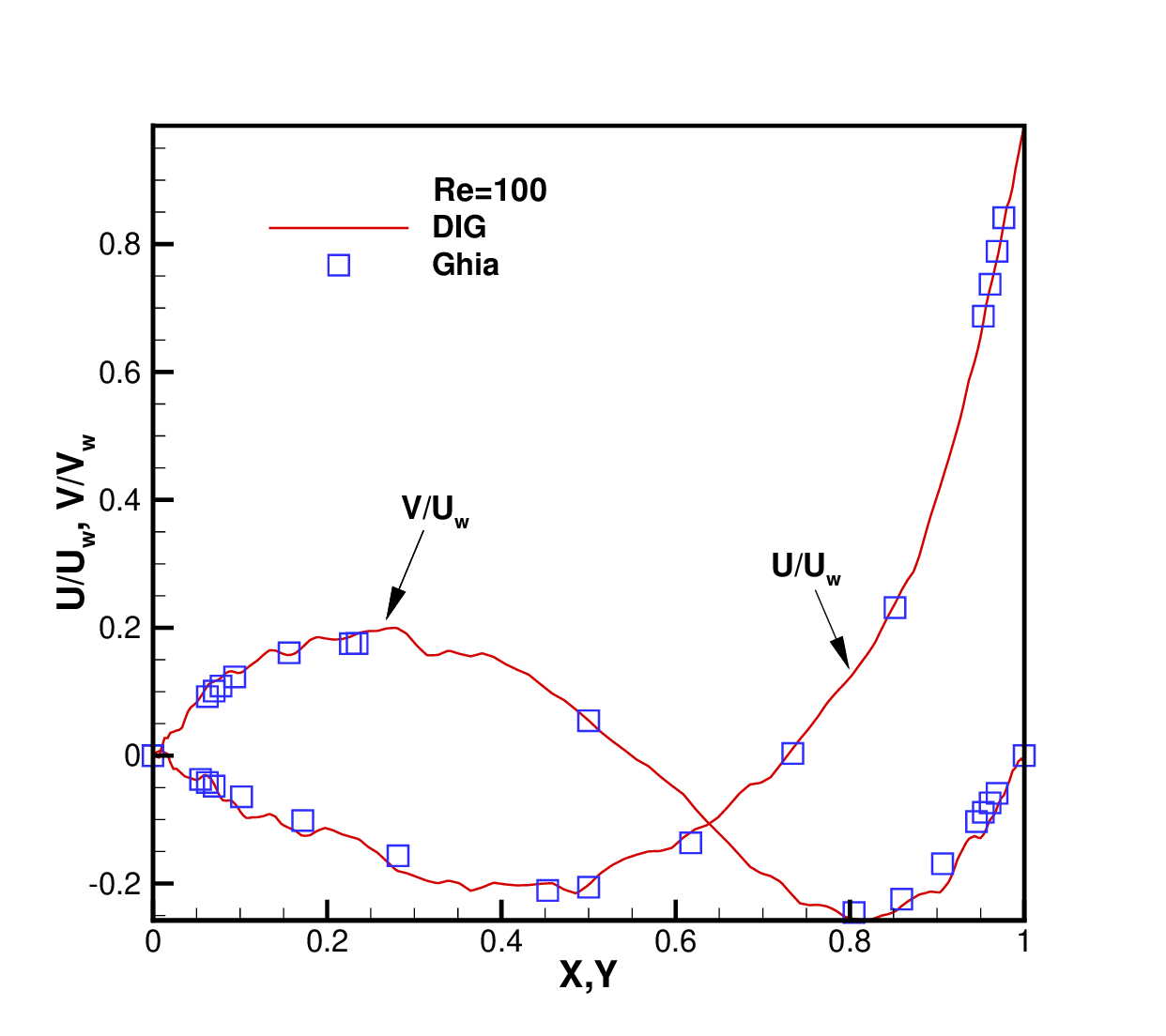}
    \includegraphics[width=0.49\textwidth]{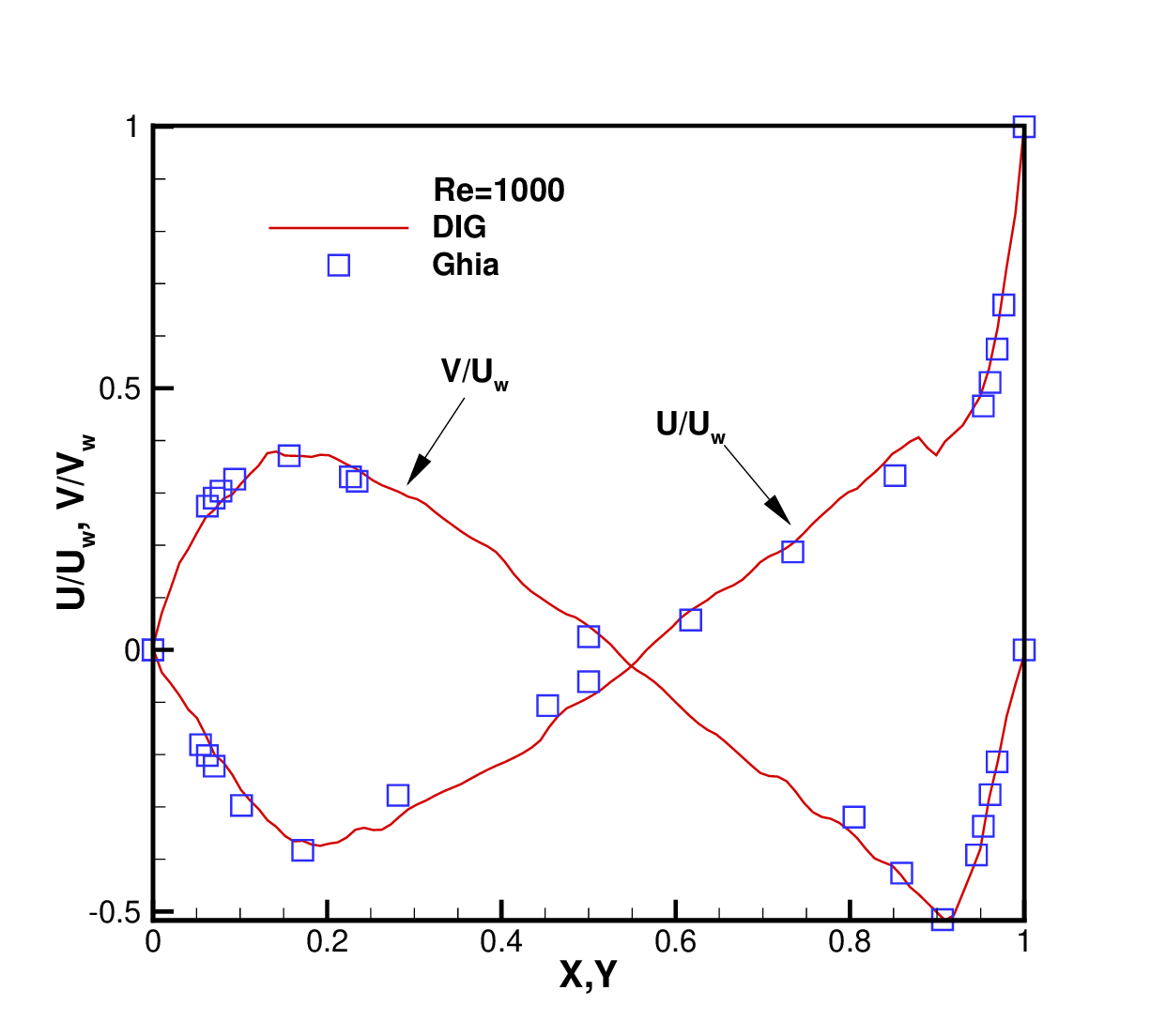}
    \caption{
    Temperature contours and heat flux streamlines in the lid-driven cavity flow, when Kn=0.1 (first row) and 0.01 (second row). For temperature contours, the colored backgrounds (black lines) represent the results obtained by DIG (DSMC), while for heat flux streamlines, DIG (DSMC) results are denoted by red (black) lines.
    (Third row) Horizontal ($U$, at $x=0.5$) and vertical ($V$, at $y=0.5$) velocities in the lid-driven flow with the Reynolds number being 100 and 1000 (or $\Kn=2.63\times10^{-3}$ and $5.26\times10^{-4}$). Ghia's data of Navier-Stokes equations are obtained from Ref.~\cite{Ghia1982}. 
    }
    \label{fig:cavitycontour}
\end{figure}


Figure~\ref{fig:cavitycontour} shows the temperature contour and streamlines of heat flux, and good agreement are observed between the DSMC and DIG  when $\Kn=0.1$ and 0.01. The DIG is applied further into the continuum flow regime, when the Knudsen numbers are $\Kn=2.63\times10^{-3}$ and $5.26\times10^{-4}$, and the Navier-Stokes equation can be used to describe the gas dynamics. The third row of Fig.~\ref{fig:cavitycontour} depicts the velocity profiles on the vertical and horizontal central lines of the cavity, and good agreements between the DIG and Ghia's benchmark solutions~\cite{Ghia1982} are observed.


Therefore, the fast convergence and asymptotic preservation of the DIG method are also notably evident in the lid-driven cavity flow. Consequently, the superiority of DIG in comparison to the DSMC is distinctly observed in Table \ref{tab2}, particularly when the Knudsen number is less than 0.1.

\section{Conclusions and outlooks}\label{sec_conclusion}

We have developed a direct intermittent GSIS-DSMC solver (DIG) to simulate the rarefied gas dynamics. The solver inherits the fast convergence and asymptotic-preserving properties of the GSIS, since the macroscopic synthetic equation is constructed and solved exactly as the GSIS in deterministic solver~\cite{Zhu2021JCP,Liu2024JCP}. However, in order to strike a balance between the computational efficiency and accuracy in the stochastic solver, the synthetic equation is solved (and its solutions of macroscopic conservative quantities are only fed back to the standard DSMC) intermittently, in every 50 simulation steps of DSMC. Such a feedback is achieved by a simple linear transformation of particles' velocities, which satisfies the principles of mass, momentum, and energy conservation. 
Numerical simulations of the hypersonic flow passing over a cylinder and the lid-driven cavity flow have demonstrated DIG's efficiency and accuracy. 


It is noted that, although the restriction on spatial resolution is removed in DIG, that on the time step is not. In fact, in our simulations the effective time step is roughly the same as the mean collision time~\footnote{From the physical perspective, this small time step is necessary, since in order to capture the Knudsen layer near the solid surface, the spatial cell size should be smaller than the mean free path, and hence the time step in DSMC is smaller than the mean collision time.}. If much larger time step is used, applying the synthetic equation every 100 steps will lead to numerical errors, as demonstrated in our preliminary tests in Fig.~\ref{Poiseuille_test}. However, the disadvantage of a smaller time step, which leads to a slower evolution, is tackled by the synthetic equation, which swiftly steers the DSMC towards steady state. 

It is also noted that, compared to the hybrid Navier-Stokes-DSMC method~\cite{Wang2003}, the DIG solves the macroscopic synthetic equation, which is derived exactly from the Boltzmann equation, in the whole computational domain. Therefore, no empirical parameters are needed to designate the continuum flow regime for applying the Navier-Stokes solver. 
The DIG is also different to the moment guided Monte Carlo method~\cite{Degond2011}, where the macroscopic equations, without the explicitly inclusion of Newton's law of viscosity and Fourier's law of heat conduction, are solved explicitly using the same time step as the DSMC. As a consequence, unlike the DIG, the flow information is not adequately exchanged. If, however, a larger time step is applied to solve the macroscopic equations in the moment guided Monte Carlo method, the macroscopic equation would become unstable~\cite{Zeng2023CiCP}. 

Given the minimal adjustments that the GSIS makes to the standard DSMC, the DIG is poised to substantially accelerate simulations of polyatomic gas flow, gas mixture flow, and even hypersonic flows with complicated chemical reactions. Also, the DIG can be applied to improve the efficiency of the PIC-MCC (particle-in-cell, Monte Carlo collision) method for plasma simulations, where species of disparate masses make the traditional method extremely slow.

\appendix
\section{The synthetic equation and the macroscopic sampling}\label{appendix}

In the gas kinetic theory, macroscopic flow quantities are obtained by taking the moments of the velocity distribution function $f$: 
\begin{equation}
\begin{aligned}
&\rho = \int_{\mathbb{R}^3}fd\bm{v},\qquad\bm{u} = \frac{1}{\rho}\int_{\mathbb{R}^3}\bm{v}fd\bm{v},\qquad T = \frac{1}{3\rho}\int_{\mathbb{R}^3}\bm{c}^2fd\bm{v}, \\
&\sigma_{ij}=\int_{\mathbb{R}^3}c_{\langle i}c_{j\rangle}fd\bm{v}, \qquad
\bm{q}=\frac{1}{2}\int_{\mathbb{R}^3}\bm{c}\bm{c}^2fd\bm{v},
\end{aligned}
\label{eq:rhout}
\end{equation}
where $\bm{c}=\bm{v}-\bm{u}$ is the peculiar velocity, and $c_{\langle i}c_{j\rangle}$ is a trace-less tensor.
Note that all variables are written in their dimensionless forms related to reference length $L$, reference density $\rho_0$, reference temperature $T_0$, and most probable speed $c_0=\sqrt{k_BT_0/m}$, where $k_B$ and $m$ are respectively the Boltzmann constant and molecular mass. The stress $\bm{\sigma}$ and heat flux $\bm{q}$ are respectively normalized by $\rho c_0^2$ and $\rho c_0^3$.

The evolution of the density, velocity and temperature is governed by the following synthetic equation:
    \begin{equation}
    \begin{aligned}
    \frac{\partial \rho}{\partial t}+\nabla\cdot(\rho \bm{u})=0&, \\
    \frac{\partial \rho\bm{u}}{\partial t}+\nabla\cdot(\rho\bm{u}\bm{u})+\nabla p+\nabla\cdot\bm{\sigma}=0&,\\
    \frac{\partial \rho E}{\partial t}+\nabla\cdot\left(\rho E\bm{u}+p\bm{u}+\bm{u}\cdot\bm{\sigma}+\bm{q}\right)=0&,
    \end{aligned}
    \label{eq:Navior-Stokes}
    \end{equation}
    where $p=\rho T$ and $E=\frac{3}{2}\rho T+\frac{1}{2}\rho u^2$. In general rarefied gas flows, the stress and heat flux cannot be expressed in terms of the velocity and temperature gradients only. Therefore, In GSIS, the constitutive relations are decomposed into two parts~\cite{Zhu2021JCP,Luo2024arXiv}:
    \begin{equation}
    \begin{aligned}
    \sigma_{ij}&=\sigma_{ij,\text{NS}}+\text{HoT}_{\sigma_{ij}},\quad 
    \bm{q}=\bm{q}_{\text{NS}}+\text{HoT}_{\bm{q}},
    \end{aligned}
    \label{eq:full_stress_heatflux}
    \end{equation}
    where first part describes the Newton law of viscosity and Fourier law of heat conduction ($\delta_{ij}$ is the Kronecker delta):
\begin{equation}
  \sigma_{ij,\text{NS}} = -\mu\left(\frac{\partial u_i}{\partial x_j}+\frac{\partial u_j}{\partial x_i}-\frac{2}{3}\delta_{ij}\nabla\cdot\bm{u}\right),\quad
\bm{q}_{\text{NS}}=-\kappa \nabla T, 
\end{equation}
and the second part describes the rarefaction effects (high-order terms, HoTs):
\begin{equation}\label{HoTs}
   \text{HoT}_{\sigma_{ij}}=\int f^*c_{\langle i}^*c_{j\rangle}^*d\bm{v}-\sigma_{ij,NS}^*,\quad
   \text{HoT}_{\bm{q}}=\frac{1}{2}\int f^*\bm{c}^*\left(c^*\right)^2d\bm{v}-\bm{q}_{\text{NS}}^*.
\end{equation}
Note that variables marked by * are obtained from the DSMC. When solving the synthetic equation to the steady state, HoTs are fixed, and therefore, the updated $\rho$, $\bm{u}$ and $T$ are different to those in Eq.~\eqref{HoTs}, during the transition state. However, when the steady state is reach, there will be little difference between $\sigma_{ij,\text{NS}}$ and $\sigma_{ij,NS}^*$ (and so is for the heat flux). Therefore, the synthetic equation~\eqref{eq:Navior-Stokes} can be viewed as exactly derived from the Boltzmann equation (equivalently, DSMC), which applies to the continuum, slip, transition, and free-molecular flow regimes.

Now we discuss how to obtain the macroscopic quantities in DSMC to be utilized in the synthetic equation. According to Eq.~\eqref{eq:rhout}, macroscopic properties at the $k-$th time step in a computational cell with volume $V_\text{cell}$ and number of particles $N_p$ can be calculated as:
\begin{equation}
\begin{aligned}
&\rho=\frac{N_\text{eff}}{V_\text{cell} }N_{p},\,\, u_i=\frac{1}{N_p}\sum_{p=1}^{N_{p}}v_{i},\,\, T=\frac{1}{3N_p}\sum_{p=1}^{N_{p}}\left|\bm{v}-\bm{u}\right|^2,\\
&\sigma_{ij}= \frac{N_\text{eff}}{V_\text{cell} }\sum_{p=1}^{N_p}\left[\left(v_{i}-u_i\right)\left(v_{j}-u_j\right)-\frac{\delta_{ij}}{3}\left|\bm{v}-\bm{u}\right|^2\right],\\
&q_i=\frac{N_\text{eff}}{2V_\text{cell} }\sum_{p=1}^{N_p}\left(v_{i}-u_i\right)\left|\bm{v}-\bm{u}\right|^2,
\end{aligned}
\label{eq:statisticmacro}
\end{equation}
where the parameter $N_\text{eff}$ is the number of real molecules represented by one simulated particle. Usually, the number of simulation particles in each cell is small, such that these quantities have significant fluctuations, and cannot be directly used in the synthetic equation. The time-averaged strategy may be used, but since in DIG the DSMC only runs a few (e.g., $N=50$) steps, the fluctuation is still large. Here, the exponentially weighted moving time averaging method~\cite{Jenny2010JCP} is employed to reduce thermal fluctuations. That is, the summation variables $\Phi=\left\{\rho, \rho\bm{u}, \rho T, \sigma_{ij}, q_i\right\}$ in Eq.~\eqref{eq:statisticmacro} are calculated as:
\begin{equation}
    \Phi(t)=\frac{n_a-1}{n_a}\Phi(t-\Delta t)+\frac{1}{n_a}\frac{N_\text{eff}}{V_\text{cell}}\sum_{p=1}^{N_p}s_k(t),
\end{equation}
where $n_a$ is the number of time steps applied in time-averaging process ($n_a=100$ in this paper), and $s_k=\left\{1, v_{i}, |\bm{c}|^2/3, c_{\langle i}c_{j\rangle}, c_{i}|\bm{c}|^2/2 \right\}$.

\vspace{1cm}

\noindent \textbf{Acknowledgments}\\
The authors thank Qi Li and Ruifeng Yuan for helpful discussions.\\

\noindent \textbf{Authors' contributions}\\
Wu and Luo contributed to conceptualization and programming, respectively. Both authors analyzed the data, wrote and approved the final manuscript.\\

\noindent \textbf{Funding}\\
This work is supported by the National Natural Science Foundation of China (12172162) and the Stable Support Plan 80000900019910072348.\\

\noindent\textbf{Availability of data and material}\\
All data generated or analyzed during this study are included in this published article.

\section*{Declarations}
\noindent\textbf{Competing interests}:
The authors declare no competing interests.

\bibliographystyle{ieeetr} 
\bibliography{thesisBib.bib}

\begin{thebibliography}{10}

\bibitem{Bird1994}
G.~A. Bird, {\em Molecular Gas Dynamics and the Direct Simulation of Gas
  Flows}.
\newblock Oxford University Press Inc, New York: Oxford Science Publications,
  1994.

\bibitem{Su2020SIAM}
W.~Su, L.~H. Zhu, and L.~Wu, ``Fast convergence and asymptotic preserving of
  the general synthetic iterative scheme,'' {\em SIAM J. Sci. Comput.},
  vol.~42, pp.~B1517--B1540, 2020.

\bibitem{Wang2003}
W.~L. Wang and I.~Boyd, ``Hybrid {DSMC}-{CFD} simulations of hypersonic flow
  over sharp and blunted bodies,'' in {\em 36th AIAA Thermophysics Conference},
  pp.~1--13, 2003.

\bibitem{Pareschi2001}
L.~Pareschi and G.~Russo, ``Time relaxed {M}onte {C}arlo methods for the
  {B}oltzmann equation,'' {\em SIAM J. Sci. Comput}, vol.~23, pp.~1253--1273,
  2001.

\bibitem{Ren2014JCP}
W.~Ren, H.~Liu, and S.~Jin, ``{An asymptotic-preserving Monte Carlo method for
  the Boltzmann equation},'' {\em J. Comput. Phys.}, vol.~276, pp.~380--404,
  2014.

\bibitem{Fei2023JCP}
F.~Fei, ``{A time-relaxed Monte Carlo method preserving the Navier-Stokes
  asymptotics},'' {\em J. Comput. Phys.}, vol.~486, p.~112128, 2023.

\bibitem{bhatnagar1954model}
P.~L. Bhatnagar, E.~P. Gross, and M.~Krook, ``A model for collision processes
  in gases. {I}. {Small} amplitude processes in charged and neutral
  one-component systems,'' {\em Phys. Rev.}, vol.~94, p.~511, 1954.

\bibitem{Shakhov_S}
E.~M. Shakhov, ``{Generalization of the Krook kinetic relaxation equation},''
  {\em Fluid Dyn.}, vol.~3, no.~5, pp.~95--96, 1968.

\bibitem{UGKS2010JCP}
K.~Xu and J.~C. Huang, ``A unified gas-kinetic scheme for continuum and
  rarefied flows,'' {\em J. Comput. Phys.}, vol.~229, pp.~7747--7764, 2010.

\bibitem{zhuyajun2016}
Y.~J. Zhu, C.~W. Zhong, and K.~Xu, ``Implicit unified gas-kinetic scheme for
  steady state solutions in all flow regimes,'' {\em J. Comput. Phys.},
  vol.~315, pp.~16--38, 2016.

\bibitem{SuArXiv2019}
W.~Su, L.~H. Zhu, P.~Wang, Y.~H. Zhang, and L.~Wu, ``Can we find steady-state
  solutions to multiscale rarefied gas flows within dozens of iterations?,''
  {\em J. Comput. Phys.}, vol.~407, p.~109245, 2020.

\bibitem{Liu2024JCP}
W.~Liu, Y.~B. Zhang, J.~N. Zeng, and L.~Wu, ``{Further acceleration of
  multiscale simulation of rarefied gas flow via a generalized boundary
  treatment},'' {\em J. Comput. Phys.}, vol.~503, p.~112830, 2024.

\bibitem{Degond2011}
P.~Degond, G.~Dimarco, and L.~Pareschi, ``{The moment guided Monte Carlo
  method},'' {\em Int. J. Numerical Methods in Fluids}, vol.~67, pp.~189--213,
  2011.

\bibitem{Liu2018arXiv}
C.~Liu, Y.~J. Zhu, and K.~Xu, ``{Unified gas-kinetic wave-particle methods I:
  Continuum and rarefied gas flow},'' {\em arXiv:1811.07141v1}, 2018.

\bibitem{Luo2024arXiv}
L.~Y. Luo, Q.~Li, F.~Fei, and L.~Wu, ``{Boosting the convergence of DSMC by
  GSIS},'' {\em arXiv:2406.16639v2}, 2024.

\bibitem{Zeng2024}
J.~N. Zeng, Q.~Li, and L.~Wu, ``General synthetic iterative scheme for rarefied
  gas mixture flows,'' {\em arXiv:2405.01099}, 2024.

\bibitem{Su2021CMAME}
W.~Su, Y.~H. Zhang, and L.~Wu, ``Multiscale simulation of molecular gas flows
  by the general synthetic iterative scheme,'' {\em Comput. Methods Appl. Mech.
  Engrg.}, vol.~373, p.~113548, 2021.

\bibitem{Zhang2023arXiv}
Y.~B. Zhang, J.~N. Zeng, R.~F. Yuan, W.~Liu, Q.~Li, and L.~Wu, ``{Efficient
  parallel solver for rarefied gas flow using GSIS},'' {\em
  arXiv:2310.18916v2}, 2023.

\bibitem{Radtke2009PRE}
G.~A. Radtke and N.~G. Hadjiconstantinou, ``{Variance-reduced particle
  simulation of the Boltzmann transport equation in the relaxation-time
  approximation},'' {\em Phys. Rev. E}, vol.~79, p.~056711, 2009.

\bibitem{Luo2023AiA}
L.~Y. Luo, Q.~Li, and L.~Wu, ``Boosting the convergence of low-variance {DSMC
  by GSIS},'' {\em Advances in Aerodynamics}, vol.~5, p.~10, 2023.

\bibitem{Wild1951}
E.~Wild, ``{On Boltzmann's equation in the kinetic theory of gases},'' {\em
  Mathematical Proceedings of the Cambridge Philosophical Society}, vol.~47,
  pp.~602--609, 1951.

\bibitem{Grad1949}
H.~Grad, ``On the kinetic theory of rarefied gases,'' {\em Comm. Pure Appl.
  Math.}, vol.~2, pp.~331--407, 1949.

\bibitem{LeiJCP2017}
L.~Wu, J.~Zhang, H.~H. Liu, Y.~H. Zhang, and J.~M. Reese, ``A fast iterative
  scheme for the linearized {Boltzmann} equation,'' {\em J. Comput. Phys.},
  vol.~338, pp.~431--451, 2017.

\bibitem{Jenny2010JCP}
P.~Jenny, M.~Torrilhon, and S.~Heinz, ``A solution algorithm for the fluid
  dynamic equations based on a stochastic model for molecular motion,'' {\em J.
  Comput. Phys.}, vol.~229, pp.~1077--1098, 2010.

\bibitem{Ghia1982}
U.~Ghia, K.~N. Ghia, and C.~T. Shin, ``{High-Re solutions for incompressible
  flow using the Navier-Stokes equations and a multigrid method},'' {\em
  Journal of Computational Physics}, vol.~48, pp.~387--411, 12 1982.

\bibitem{Zhu2021JCP}
L.~H. Zhu, X.~C. Pi, W.~Su, Z.~H. Li, Y.~H. Zhang, and L.~Wu, ``General
  synthetic iterative scheme for nonlinear gas kinetic simulation of
  multi-scale rarefied gas flows,'' {\em J. Comput. Phys.}, vol.~430,
  p.~110091, 2021.

\bibitem{Zeng2023CiCP}
J.~N. Zeng, W.~Su, and L.~Wu, ``General synthetic iterative scheme for unsteady
  rarefied gas flows,'' {\em Commun. Comput. Phys.}, vol.~34, pp.~173--207,
  2023.

\end{thebibliography}

\end{document}